\pdfoutput=1
\documentclass[final,12pt,nopreprintline,longtitle]{elsarticle}

\usepackage{amssymb}
\usepackage{amsmath}
\usepackage{lineno}
\usepackage[hidelinks]{hyperref}
\usepackage{cleveref}
\usepackage{geometry}
\usepackage{enumitem}
\usepackage{caption}
\usepackage{booktabs}
\usepackage{graphicx}
\usepackage{microtype}
\usepackage{tikz}
\usepackage{float}
\usepackage[section]{placeins}
\usepackage{paracol}

\newcommand{\paperbibliography}[1]{\bibliography{#1}}

\journal{Computer Physics Communications}

\renewcommand{\v}{\mathbf{v}}
\newcommand{\V}{\mathbf{V}}
\newcommand{\x}{\mathbf{x}}
\renewcommand{\k}{\mathbf{k}}
\newcommand{\h}{\boldsymbol{h}}

\newcommand{\Sf}{\mathbf{S}_f}

\newcommand{\betaac}{\beta}
\newcommand{\plotAidDisclosure}{Plotting code was prepared with assistance from
OpenAI Codex (model \texttt{gpt-6-astra}); see \cref{sec:aiWorkflow}.}
\newcommand{\schematicAidDisclosure}{Drawing code was prepared with assistance from
OpenAI Codex (model \texttt{gpt-6-astra}).}

\makeatletter
\def\ps@pprintTitle{%
    \let\@oddhead\@empty
    \let\@evenhead\@empty
    \let\@oddfoot\@empty
    \let\@evenfoot\@empty}
\makeatother

\begin{document}

\begin{frontmatter}

\title{An unstructured finite-volume Helmholtz method with perfectly matched layers for heterogeneous two-phase acoustics}

\author[mma]{Chuanchao Xu}
\ead{xu@mma.tu-darmstadt.de}
\author[mma]{Jun Liu}
\author[must]{Jan Helge Dörsam}
\author[must]{Mario Kupnik}
\author[mma]{Tomislav Maric\texorpdfstring{\corref{cor1}}{}}
\ead{maric@mma.tu-darmstadt.de}
\author[mma]{Dieter~Bothe}
\ead{bothe@mma.tu-darmstadt.de}
\cortext[cor1]{Corresponding author.}

\affiliation[mma]{organization={Mathematical Modeling and Analysis,
    Mathematics Department, TU Darmstadt},
    addressline={Peter-Gruenberg-Strasse 10},
    city={Darmstadt}, postcode={64287},
    country={Germany}}
\affiliation[must]{organization={Measurement and Sensor Technology,
    Department of Electrical Engineering and Information Technology,
    TU Darmstadt},
    addressline={Merckstrasse 25},
    city={Darmstadt}, postcode={64283},
    country={Germany}}

\begin{abstract}
\small
This paper presents a cell-centred unstructured finite-volume method for
time-harmonic two-phase acoustics.  A volume-fraction representation supplies
density and acoustic compressibility.  Face fractions average available
neighbouring planar interface cuts independently of velocity direction,
retaining the established one-field face-density closure.  Cartesian
complex-coordinate stretching provides perfectly matched layers (PMLs).
Real--imaginary splitting yields a real-valued block system, discretized with
corrected non-orthogonal fluxes and consistent boundary contributions.
Verification covers homogeneous waves, layered gas--liquid transmission with
PML truncation, baffled-piston radiation with Kirchhoff far-field
reconstruction, and resolved rigid-sphere radiation forces.  Homogeneous-wave
pressure converges at approximately second order on orthogonal meshes and on
meshes with non-orthogonal interiors and orthogonal boundary cells.
Reconstructed velocity converges at approximately second order on orthogonal
meshes and with an observed order of about $1.7$ on the latter mesh family.
Under aligned refinement, the layered case's relative whole-domain
complex-pressure $L_2$ error decreases to $2.657\times10^{-4}$.  The resolved-sphere
force differs from the Gorkov prediction by at most $1.5\%$ over the tested
Rayleigh size range.  The results quantify the accuracy and current limitations
of the heterogeneous Helmholtz--PML formulation.
\end{abstract}


\end{frontmatter}

\section{Introduction}
\label{sec:intro}

Time-harmonic pressure formulations avoid resolving many acoustic periods when
the response at a prescribed frequency is sought.  They are therefore widely
used for resonators, radiation problems, and transducer-driven devices.  Their
application to gas--liquid configurations is more demanding because density and
sound speed are discontinuous at the material interface.  Acoustic levitation of a
liquid drop is one motivating example: reflection and transmission at the drop
affect the pressure and momentum-flux fields that determine its equilibrium and
stability \cite{andrade2019numerical}.

Frequency-domain acoustic equations have previously been discretized by finite
volumes on unstructured meshes.  Nicoud et al.\ developed a cell-vertex
tetrahedral Helmholtz solver for thermoacoustic eigenmodes with complex
impedance and flame coupling \cite{Nicoud2007}, while Long et al.\ formulated a
three-dimensional finite-volume sound model on horizontally triangular,
vertically layered coastal-ocean grids \cite{Long2015}.  These studies
demonstrate the geometrical flexibility of finite-volume Helmholtz solvers, but
they do not combine a volume-of-fluid (VOF) gas--liquid material description
with PML truncation on a cell-centred polyhedral mesh.

Most interface-specific treatments of discontinuous acoustic coefficients use
structured grids.  Zhang and LeVeque obtained second-order accuracy for
time-domain acoustics by incorporating jump conditions into an
immersed-interface method on Cartesian grids \cite{ZhangLeVeque1997}.  Baruch
et al.\ constructed high-order one-dimensional schemes for time-harmonic
equations with discontinuous coefficients \cite{Baruch2009}.  Jeong et al.\ used
ghost points and local extensions for three-dimensional acoustic transmission
across complex interfaces on Cartesian grids \cite{Jeong2021}.  These methods
resolve material jumps explicitly, but their stencils are not directly
transferable to a cell-centred operator on arbitrary polyhedral meshes.

Geometrical VOF methods provide complementary tools for such meshes.  Planar
interface reconstruction and liquid-covered face areas can be evaluated on
general polyhedral cells, as demonstrated by isoAdvector and the reconstructed
distance-function PLIC method \cite{Roenby2016,Scheufler2019}.  Those methods
were developed for interface advection and do not by themselves provide a
discretization of the pressure Helmholtz formulation used here.

Open-domain truncation is a separate numerical requirement.  Complex-coordinate
PMLs for the time-harmonic Helmholtz equation have been analyzed in symmetric
finite-element formulations and with damping profiles designed to reduce
truncation error \cite{Harari2000,Bermudez2007}.  Stable second-order PML
formulations also exist for time-domain acoustic wave equations
\cite{Kaltenbacher2013modified}.  Finite-volume Helmholtz discretizations,
geometrical VOF representations, and PML truncation are thus established
separately.  To the authors' knowledge, no published method
combines them in a cell-centred polyhedral discretization for heterogeneous
gas--liquid acoustics.  This paper develops that integrated frequency-domain
method.

The model assumes linear, adiabatic acoustics about a quiescent base state.  The
phase interface and material properties are fixed during each frequency-domain
solve.  Material interfaces are confined to the physical domain, and each PML
contains a homogeneous continuation of the exterior medium.  Mean flow,
viscosity, heat conduction, nonlinear acoustics, and interface motion over an
acoustic period are outside the present scope.

The contribution of this work is an integrated frequency-domain formulation
that combines the heterogeneous pressure Helmholtz equation, a VOF-based
gas--liquid material representation, a cell-centred unstructured FVM, and PML
truncation.  The cell and face material coefficients are obtained from volume
and reconstructed face fractions.  A shared face fraction is formed from the
available adjacent-cell PLIC cuts, including across processor boundaries, while the
complex-coordinate PML is expressed as a coupled real-valued block system.

A companion software manuscript by Xu, Liu, Bothe and Maric is in preparation
and addresses the implementation, case workflow, parallel block assembly and
performance studies.  The present paper focuses on the numerical formulation
and its verification; the implementation used here is identified in the code
and data availability statement.

The remainder of the paper is organized as follows.  \Cref{sec:model} derives
the heterogeneous frequency-domain acoustic model, radiation-pressure
quantity, and PML formulation.  \Cref{sec:discretization} defines the two-phase
material representation and derives the unstructured FVM block discretization.
\Cref{sec:verification} presents the homogeneous plane-wave, layered
gas--liquid--PML, baffled-piston, and rigid-sphere studies.  Finally,
\Cref{sec:discussion} discusses the coefficient treatment, PML behavior, and
remaining accuracy limitations.

\section{Mathematical model}
\label{sec:model}

\subsection{Wave equation for heterogeneous media}
Although standard, the linear acoustic derivation is retained to show how the
spatially varying density enters the conservative pressure equation and to
establish the time convention used in the subsequent Helmholtz, PML, and
finite-volume formulations.
The governing equations of fluid motion for density $\rho$, pressure $p$, and velocity $\v$ are expressed as follows:
\begin{itemize}
\item Mass conservation
        \begin{equation}
            \frac{\partial \rho}{\partial t} + \nabla\cdot(\rho \v)=0.
            \label{eq:massConserv}
        \end{equation}
        \item Momentum conservation, neglecting gravity and viscosity
        \begin{equation}
            \frac{\partial \rho \v}{\partial t} + \nabla\cdot(\rho \v\v)=-\nabla p.
            \label{eq:momConserv}    
        \end{equation}
\end{itemize}  
   
\begin{samepage}
Consider a quiescent base state that is stationary but spatially heterogeneous,
with $\v_0=\mathbf{0}$, $\rho_0=\rho_0(\x)$,
$c=c(\x)$, and $p_0=p_0(\x)$.  All base-state fields are independent of time.
Classical perturbation theory represents the acoustic field by the first-order
quantities $\rho_1$, $p_1$, and $\v_1$, with
$|\rho_1|\ll\rho_0$ and $|p_1|\ll|p_0|$.  The fields are decomposed as
    \begin{equation}
        \begin{aligned}
            \rho(\x,t) &= \rho_0(\x) + \rho_1(\x,t), \\
            p(\x,t) &= p_0(\x) + p_1(\x,t), \\
            \v(\x,t) &= \v_1(\x,t).
            \label{eq:1stPert}
        \end{aligned}
    \end{equation}
\end{samepage}

At zeroth order, \cref{eq:momConserv} gives $\nabla p_0=0$ within each
bulk region.  Thus, $p_0$ is constant there, whereas the material fields
$\rho_0(\x)$ and $c(\x)$ may still vary spatially.
    
Substituting \cref{eq:1stPert} into \cref{eq:momConserv} and neglecting higher-order terms (e.g., $\rho_1\v_1$, $\v_1\v_1$) yields
        \begin{equation}
            \rho_0 \frac{\partial \v_1}{\partial t}=-\nabla p_1.
            \label{eq:1stMomConserv}            
        \end{equation}

After introducing the total derivative $\frac{d}{dt}:=\frac{\partial}{\partial t} + \v \cdot \nabla$, \cref{eq:massConserv} can be written as
\begin{equation}
    \frac{d \rho}{dt} + \rho \nabla\cdot\v =0.
\end{equation}

Sound propagation is modelled as adiabatic, so $\frac{d s}{d t}=0$, where $s$
is the specific entropy, and we have \cite{LandauLifshitz1987}
\begin{equation}
     \frac{d \rho}{dt}
     =\left(\frac{\partial \rho}{\partial p}\right)_s\frac{d p}{d t}
     +\left(\frac{\partial \rho}{\partial s}\right)_p\frac{d s}{d t}
     =\frac{1}{c^2}\frac{d p}{d t}
     =\frac{1}{c^2}\left(\frac{\partial p}{\partial t}+(\v\cdot\nabla)p\right),
\end{equation}
where we have used the isentropic sound speed
$c:=\sqrt{(\partial p/\partial \rho)_s}$, evaluated locally at the base state.
The mass conservation equation
now becomes
\begin{equation}
    \frac{\partial p}{\partial t} + (\v\cdot\nabla)p + \rho c^2 \, \nabla\cdot\v = 0.
    \label{eq:pMassConti}
\end{equation}

Following the same procedure, we use the first-order perturbation ansatz \cref{eq:1stPert}. Neglecting higher-order terms gives
\begin{equation}
    \frac{\partial p_1}{\partial t} = -\rho_0 c^2\, \nabla \cdot \v_1.
    \label{eq:1stMassConserv}
\end{equation}
Taking the time derivative and then using \cref{eq:1stMomConserv} to eliminate $\v_1$ gives
\begin{equation}
    \frac{1}{\rho_0(\x)c^2(\x)}
    \frac{\partial^2p_1}{\partial t^2}
    -\nabla\cdot\left(\frac{1}{\rho_0(\x)}\nabla p_1\right)=0,
    \label{eq:heteWaveP}
\end{equation}
which is the pressure wave equation for a heterogeneous medium.  Both
coefficients retain their spatial dependence.  Taking $\rho_0$ and $c$ as
constants gives the familiar homogeneous wave equation
\begin{equation}
    \frac{\partial^2 p_1}{\partial t^2} = c^2 \Delta p_1.
\end{equation}

\subsection{Helmholtz equation}
For a linear acoustic system driven at a single angular frequency, the
steady response after transients have decayed can be written in the following
time-harmonic form \cite{morse1986theoretical}
\begin{equation}
    \begin{aligned}
        p_1(\x,t) &= \mathrm{Re}(e^{-i\omega t}P(\x)),\\
        \v_1(\x,t) &= \mathrm{Re}(e^{-i\omega t}\V(\x)).
    \end{aligned}
    \label{as:timeHarmonic}
\end{equation}
where the complex pressure
$P(\x)=P_{re}(\x)+iP_{im}(\x)$ is a spatial function.
Inserting \cref{as:timeHarmonic} into the first-order momentum
equation~\eqref{eq:1stMomConserv} gives
\begin{equation}
    i\omega\rho_0 \V = \nabla P,
    \label{eq:velocityComplex}
\end{equation}
or equivalently $\V=\nabla P/(i\omega\rho_0)$.  Thus,
$\V_{re}=\nabla P_{im}/(\omega\rho_0)$ and
$\V_{im}=-\nabla P_{re}/(\omega\rho_0)$.
Substituting \cref{as:timeHarmonic} into the general wave
equation~\eqref{eq:heteWaveP} then gives
\begin{equation}
    \nabla\cdot\bigl(a(\x)\nabla P\bigr)
    +\omega^2\betaac(\x)P=0,
    \qquad
    a(\x):=\frac{1}{\rho_0(\x)},\quad
    \betaac(\x):=\frac{1}{\rho_0(\x)c^2(\x)}.
    \label{eq:heteHelmholtz}
\end{equation}
Equivalently, multiplication by $\rho_0$ gives
$\rho_0\nabla\cdot(\nabla P/\rho_0)+k^2P=0$, where the local wavenumber is
$k(\x)=\omega/c(\x)$.

\paragraph{Sharp two-phase special case}
Equation~\eqref{eq:heteHelmholtz} describes a generally heterogeneous medium.
For a sharp gas--liquid interface $\Gamma$, its conservative form is
interpreted weakly over the complete domain, with the piecewise-constant
coefficients
\begin{equation}
    \rho_0=\chi_l\rho_l+(1-\chi_l)\rho_g,
    \qquad
    a:=\frac{1}{\rho_0},
    \qquad
    \betaac:=\frac{1}{\rho_0c^2}
      =\frac{\chi_l}{\rho_lc_l^2}
       +\frac{1-\chi_l}{\rho_gc_g^2},
    \label{eq:sharpTwoPhaseCoefficients}
\end{equation}
where $\chi_l$ is the liquid indicator, equal to one in the liquid and zero in
the gas.  The corresponding conservative Helmholtz equation is
$\nabla\cdot(a\nabla P)+\omega^2\betaac P=0$.  The equilibrium interface
geometry is fixed during each solve, and first-order capillary--acoustic
coupling is neglected.  This is appropriate away from capillary resonances
when the resulting interfacial stress is small relative to the acoustic stress
\cite{Krechetnikov2019,ShapiroWeinstein2011}; for the planar one-dimensional
interface considered below, the curvature perturbation vanishes identically
\cite{Krechetnikov2019}.  Pressure and its normal flux then satisfy
\begin{equation}
    [P]_\Gamma=0,
    \qquad
    [a\nabla P\cdot\mathbf n_\Gamma]_\Gamma=0,
    \label{eq:interfaceJumpConditions}
\end{equation}
where $[\phi]_\Gamma=\phi_l-\phi_g$ at the interface.  Surface tension therefore determines the
equilibrium pressure jump of the base state but does not enter the present
first-order acoustic problem.  \Cref{sec:materialRepresentation} introduces
the VOF representation used to discretize these discontinuous coefficients.

\subsection{Acoustic radiation pressure}
The time-averaged force on a droplet in an acoustic levitator is obtained from
the acoustic radiation pressure. Its relation to the computed first-order
acoustic fields is summarized below.
The following derivation assumes a homogeneous medium, so the unperturbed
density $\rho_0$ and sound speed $c$ are constants rather than spatial fields.
The fields are expanded to second order as
\begin{equation}
    \begin{aligned}
        \rho(\x,t) &= \rho_0 + \rho_1(\x,t) + \rho_2(\x,t), \\
        p(\x,t) &= p_0 + p_1(\x,t) + p_2(\x,t), \\
        \v(\x,t) &= \v_1(\x,t) + \v_2(\x,t).
        \label{eq:2stPert}
    \end{aligned}
\end{equation}
Substituting this expansion into the momentum equation~\eqref{eq:momConserv},
neglecting terms above second order, and using the first-order momentum
equation~\eqref{eq:1stMomConserv} gives
\begin{equation}
    \nabla p_2 =
    -\rho_0\frac{\partial \v_2}{\partial t}
    -\frac{\partial (\rho_1\v_1)}{\partial t}
    -\rho_0\nabla\cdot(\v_1\v_1).
\end{equation}
Expanding $\partial(\rho_1\v_1)/\partial t$ and using the first-order
momentum and mass equations~\eqref{eq:1stMomConserv} and
\eqref{eq:1stMassConserv} yields
\begin{equation}
    \nabla p_2 =
    -\rho_0\frac{\partial \v_2}{\partial t}
    +\frac{\nabla p_1^2}{2c^2\rho_0}
    -\rho_0(\nabla\v_1)\cdot\v_1.
\end{equation}
Here the isentropic relation $p_1=c^2\rho_1$ has been used. For an
irrotational first-order velocity field, $\v_1=\nabla\Phi$, the identity
$(\nabla\v_1)\cdot\v_1=\frac{1}{2}\nabla|\v_1|^2$ applies. Moreover, the
time average of $\partial\v_2/\partial t$ over one driving period vanishes.
Denoting the time average by $\langle\cdot\rangle:= \frac{1}{T}\int_0^T \cdot \, dt$ therefore gives
\begin{equation}
    \langle p_2\rangle =
    \frac{\langle p_1^2\rangle}{2\rho_0c^2}
    -\frac{\rho_0}{2}\langle|\v_1|^2\rangle.
\end{equation}
Finally, applying the time-harmonic ansatz \cref{as:timeHarmonic} gives
\begin{equation}
    \langle p_2\rangle =
    \frac{\langle \mathrm{Re}(e^{-i\omega t}P(x))^2\rangle}{2\rho_0c^2}
    -\frac{\rho_0}{2}\langle|\mathrm{Re}(e^{-i\omega t}\V(x))|^2\rangle =  \frac{ |P(x)|^2}{4\rho_0c^2}
    -\frac{\rho_0}{4}|\V(x)|^2,
    \label{eq:acousticRadiationPressure}
\end{equation}
where the time-dependent terms vanish after time averaging.  Thus, the acoustic
radiation pressure $p_{\mathrm{rad}}:=\langle p_2\rangle$
can be evaluated from the first-order pressure and velocity fields.

\subsection{Perfectly matched layer}
As the computational domain must be bounded, open acoustic systems require a
domain-truncation treatment.  An absorbing or radiation boundary condition can
be effective for a known outgoing-wave structure, whereas a perfectly matched
layer (PML) is more suitable for general multidimensional radiation problems.
In the continuous formulation, a wave enters the PML without reflection and is
attenuated within the layer.  A finite, discretized PML can still produce a
small truncation error, whose parameter and resolution dependence is assessed in
\cref{sec:pmlSensitivity}.

Following the modified PML formulation of Kaltenbacher et al.
\cite{Kaltenbacher2013modified}, each real coordinate $x_j$ is extended into a
complex coordinate $\tilde{x}_j$.  For the time convention in
\cref{as:timeHarmonic}, the directional stretching factors are
\begin{equation}
    \eta_j
    =\frac{\partial\tilde{x}_j}{\partial x_j}
    =1+\frac{\sigma_j}{i\omega},
    \qquad
    \frac{\partial}{\partial\tilde{x}_j}
    =\frac{1}{\eta_j}\frac{\partial}{\partial x_j},
    \qquad j\in\{x,y,z\}.
    \label{eq:dampEtaDef}
\end{equation}
Here, $\sigma_j=0$ in the physical domain and $\sigma_j<0$ in the PML.  The
negative sign follows from the adopted $e^{-i\omega t}$ convention.  Equivalently,
$\tilde{\x}=\x+i\h(\x)$, with
$h_j(x_j)=-\omega^{-1}\int\sigma_j(x_j)\,dx_j$.  A plane wave then satisfies
\begin{equation}
    P=|P|\exp(i\k\cdot\tilde{\x})
     =|P|\exp(-\k\cdot\h)\exp(i\k\cdot\x),
    \label{eq:planeWaveDamp}
\end{equation}
so that $\k\cdot\h>0$ produces attenuation in the PML.

Applying \cref{eq:dampEtaDef} to the heterogeneous Helmholtz equation gives
\begin{equation}
    -\frac{(i\omega)^2}{c^2}P
    +\rho_0\sum_{j\in\{x,y,z\}}
    \frac{1}{\eta_j}\frac{\partial}{\partial x_j}
    \left(
        \frac{1}{\rho_0\eta_j}
        \frac{\partial P}{\partial x_j}
    \right)=0.
    \label{eq:extendedStretchHelmholz}
\end{equation}
For coordinate-wise separable stretching, meaning that each factor depends
only on its corresponding coordinate,
$\eta_x=\eta_x(x)$, $\eta_y=\eta_y(y)$, and $\eta_z=\eta_z(z)$.  Defining
\begin{equation}
    J=\eta_x\eta_y\eta_z,
    \qquad
    q_x=\frac{\eta_y\eta_z}{\eta_x},\quad
    q_y=\frac{\eta_x\eta_z}{\eta_y},\quad
    q_z=\frac{\eta_x\eta_y}{\eta_z},
    \label{eq:pmlJacobianRatios}
\end{equation}
and multiplying \cref{eq:extendedStretchHelmholz} by $J/\rho_0$ yields
\begin{equation}
    -\frac{J(i\omega)^2}{\rho_0c^2}P
    +\sum_{j\in\{x,y,z\}}
    \frac{\partial}{\partial x_j}
    \left(
        \frac{q_j}{\rho_0}
        \frac{\partial P}{\partial x_j}
    \right)=0.
    \label{eq:reformedStretchHelmholz}
\end{equation}

Material interfaces are confined to the physical solution domain.  Each PML
region is filled with the same homogeneous exterior medium present at its inner
boundary, and therefore $\rho_0$ and $c$ are constant throughout that PML
region.  Multiplication of \cref{eq:reformedStretchHelmholz} by this constant
$\rho_0$ consequently cancels the density from the PML flux.  This is why the
PML correction is written as $\nabla\cdot(\mathbf{T}_i\nabla P)$ rather than
$\rho_0\nabla\cdot(\mathbf{T}_i\nabla P/\rho_0)$.  In the physical domain,
where the material properties may be heterogeneous, $\boldsymbol{\sigma}=0$
and all PML correction coefficients vanish; the original heterogeneous
operator is retained there.

The mass and directional stretching factors can be separated into real and
imaginary parts as
\begin{equation}
\begin{aligned}
    (i\omega)^2J
    ={}&-\omega^2
       +(\sigma_x\sigma_y+\sigma_x\sigma_z+\sigma_y\sigma_z)\\
      &+i\left[
        \omega(\sigma_x+\sigma_y+\sigma_z)
        -\frac{\sigma_x\sigma_y\sigma_z}{\omega}
      \right],\\
    q_j={}&1+T_{0jj}-iT_{1jj}.
\end{aligned}
\label{eq:pmlComplexSplit}
\end{equation}
With $P=P_{re}+iP_{im}$, the coupled real-valued PML equations are therefore
\begin{equation}
\begin{aligned}
    \frac{\omega^2}{c^2}P_{re}
    +\rho_0\nabla\cdot\left(\frac{1}{\rho_0}\nabla P_{re}\right)
    -C_0P_{re}+C_1P_{im}
    +\nabla\cdot\left(\mathbf{T}_0\cdot\nabla P_{re}
                     +\mathbf{T}_1\cdot\nabla P_{im}\right)&=0,\\
    \frac{\omega^2}{c^2}P_{im}
    +\rho_0\nabla\cdot\left(\frac{1}{\rho_0}\nabla P_{im}\right)
    -C_0P_{im}-C_1P_{re}
    +\nabla\cdot\left(\mathbf{T}_0\cdot\nabla P_{im}
                     -\mathbf{T}_1\cdot\nabla P_{re}\right)&=0,
\end{aligned}
\label{eq:helmholtzPML}
\end{equation}
where $C_0$, $C_1$ and the tensor coefficients $\mathbf{T}_0$, $\mathbf{T}_1$
are defined by
\begin{paracol}{2}
      \begin{equation*}
        \begin{aligned}
            C_0 &= \frac{\sigma_x\sigma_y + \sigma_x\sigma_z + \sigma_y\sigma_z}{c^2} \\
            T_{0xx} &= \frac{\sigma_x(\sigma_y+\sigma_z-\sigma_x) - \sigma_y\sigma_z}{(\omega^2 + \sigma^2_x)}, \\
            T_{0yy} &= \frac{\sigma_y(\sigma_x+\sigma_z-\sigma_y)-\sigma_x\sigma_z}{(\omega^2 + \sigma^2_y)}, \\
            T_{0zz} &= \frac{\sigma_z(\sigma_x+\sigma_y-\sigma_z)-\sigma_x\sigma_y}{(\omega^2 + \sigma^2_z)}, \\
            \mathbf{T_0} &= \begin{pmatrix}
              T_{0xx} & 0 & 0 \\
             0 &  T_{0yy} & 0 \\
             0 & 0 &  T_{0zz}
            \end{pmatrix},
        \end{aligned}
    \end{equation*}
    \switchcolumn
       \begin{equation*}
        \begin{aligned}
            C_1 &= \frac{\omega(\sigma_x+\sigma_y+\sigma_z) - (\sigma_x\sigma_y\sigma_z)/\omega}{c^2} \\
            T_{1xx} &= \frac{(\sigma_x\sigma_y\sigma_z)/\omega + \omega(\sigma_y+\sigma_z-\sigma_x)}{(\omega^2 + \sigma^2_x)}, \\
            T_{1yy} &= \frac{(\sigma_x\sigma_y\sigma_z)/\omega + \omega(\sigma_x+\sigma_z-\sigma_y)}{(\omega^2 + \sigma^2_y)}, \\
            T_{1zz} &= \frac{(\sigma_x\sigma_y\sigma_z)/\omega + \omega(\sigma_x+\sigma_y-\sigma_z)}{(\omega^2 + \sigma^2_z)}, \\
            \mathbf{T_1} &= \begin{pmatrix}
              T_{1xx} & 0 & 0 \\
             0 &  T_{1yy} & 0 \\
             0 & 0 &  T_{1zz}
            \end{pmatrix}.
        \end{aligned}
    \end{equation*}
\end{paracol}

\section{Unstructured FVM Discretization}
\label{sec:discretization}

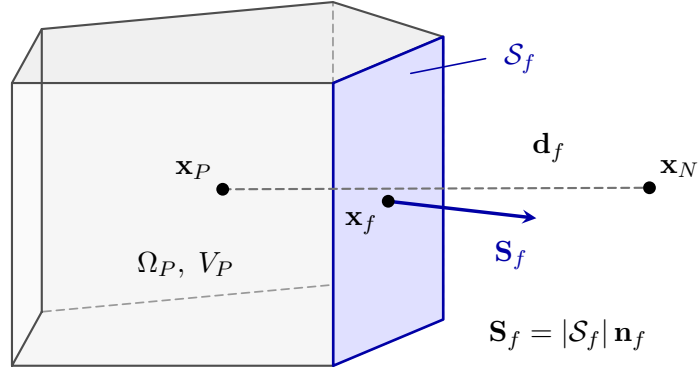
\begin{figure}[htbp]
    \centering
    \begin{tikzpicture}[
    x={(1.7cm,0cm)}, y={(0.646cm,0.51cm)}, z={(0cm,1.7cm)},
    line join=round, line cap=round, >=stealth,
    every node/.style={font=\small},
    edge/.style={draw=black!70,line width=0.8pt},
    hidden/.style={draw=black!40,densely dashed,line width=0.65pt},
    centre/.style={circle,fill=black,inner sep=1.7pt}
]
    \coordinate (A) at (0,0,0);
    \coordinate (B) at (2.5,0,0);
    \coordinate (C) at (2.9,1.2,0);
    \coordinate (D) at (1.7,2.1,0);
    \coordinate (E) at (-0.3,1.4,0);
    \coordinate (At) at (0,0,2.2);
    \coordinate (Bt) at (2.5,0,2.2);
    \coordinate (Ct) at (2.9,1.2,2.2);
    \coordinate (Dt) at (1.7,2.1,2.2);
    \coordinate (Et) at (-0.3,1.4,2.2);

    \fill[gray!7] (A)--(B)--(Bt)--(At)--cycle;
    \fill[gray!4] (E)--(A)--(At)--(Et)--cycle;
    \fill[gray!12] (At)--(Bt)--(Ct)--(Dt)--(Et)--cycle;
    \draw[hidden] (B)--(C)--(D)--(E);
    \draw[hidden] (C)--(Ct) (D)--(Dt);
    \draw[edge] (E)--(A)--(B) (E)--(Et) (A)--(At) (B)--(Bt);
    \draw[edge] (At)--(Bt)--(Ct)--(Dt)--(Et)--cycle;

    \path[fill=blue!12,draw=blue!65!black,line width=1.0pt]
        (B)--(C)--(Ct)--(Bt)--cycle;

    \coordinate (P) at (1.2937044400,0.9111994698,1.1);
    \coordinate (F) at (2.7,0.6,1.1);
    \coordinate (N) at (4.6,0.95,1.1);
    \coordinate (S) at (4.02,0.16,1.1);

    \draw[black!55,densely dashed,line width=0.8pt] (P)--(N);
    \node[fill=white,inner sep=1pt] at (3.8,1.0,1.40)
        {$\mathbf d_f$};
    \draw[->,blue!65!black,line width=1.3pt] (F)--(S)
        node[pos=0.83,below=5pt] {$\mathbf S_f$};

    \node[centre,label={[inner sep=2pt]above left:$\mathbf x_P$}] at (P) {};
    \node[centre,label={[inner sep=2pt]below left:$\mathbf x_f$}] at (F) {};
    \node[centre,label={[inner sep=2pt]above right:$\mathbf x_N$}] at (N) {};
    \node[fill=gray!7,inner sep=1.5pt] at (1.18,0.45,0.65) {$\Omega_P,\;V_P$};
    \node[anchor=west,text=blue!65!black] at (3.48,0.7,2.18)
        {$\mathcal S_f$};
    \draw[blue!65!black,line width=0.6pt]
        (3.42,0.7,2.18)--(2.78,0.85,2.02);

    \node[anchor=west] at (3.55,0.2,0.18)
        {$\mathbf S_f=|\mathcal S_f|\,\mathbf n_f$};
\end{tikzpicture}
    \caption{Polyhedral control-volume geometry for the cell-centred FVM.
    The owner cell $\Omega_P$ has volume $V_P$ and centroid $\x_P$.
    The highlighted face $\mathcal S_f$ has centroid $\x_f$ and outward
    area vector $\Sf=|\mathcal S_f|\mathbf n_f$.  The dashed line joins
    the owner centroid to the centroid $\x_N$ of an adjacent cell, whose
    boundary is omitted for clarity.  This is a schematic projection;
    $\mathbf d_f=\x_N-\x_P$ need not be parallel to $\Sf$.
    OpenAI Codex (OpenAI, model \texttt{gpt-6-astra}) assisted with generating
    and revising the TikZ drawing code.}
    \label{fig:standardFVMControlVolume}
\end{figure}

The coupled equations in \cref{eq:helmholtzPML} are discretized with a
cell-centred unstructured FVM on a polyhedral mesh
\cite{Jasak1996,Moukalled2016}.  A control volume $\Omega_P$ has volume $V_P$,
centroid $\x_P$, and boundary faces $\mathcal{S}_f$.  The outward face-area
vector is $\Sf=\mathbf{n}_f|\mathcal{S}_f|$.  The standard cell-centred
control-volume geometry is summarized in
\cref{fig:standardFVMControlVolume}.

\begin{figure}[t]
    \centering
    \resizebox{0.98\textwidth}{!}{%
    \begin{tikzpicture}[
        x=1cm,
        y=1cm,
        line cap=round,
        line join=round,
        >=stealth,
        every node/.style={font=\small},
        centre/.style={circle,fill=black,inner sep=1.8pt},
        cell/.style={draw=black!85,line width=0.9pt,fill=gray!5}
    ]
        \begin{scope}
            \node[font=\small\bfseries] at (2.8,3.85)
                {(a) Interface phase fractions};

            \path[cell]
                (-0.15,0.25)--(2.40,0.15)--(3.25,3.00)--
                (1.45,3.35)--(-0.35,2.25)--cycle;
            \path[cell]
                (2.40,0.15)--(5.50,0.10)--(6.00,1.65)--
                (5.15,3.40)--(3.25,3.00)--cycle;

            \begin{scope}
                \clip
                    (-0.15,0.25)--(2.40,0.15)--(3.25,3.00)--
                    (1.45,3.35)--(-0.35,2.25)--cycle;
                \fill[cyan!18]
                    (-0.6,-0.2)--(6.2,-0.2)--(6.2,1.42)--
                    (2.68,1.10)--(-0.6,0.80)--cycle;
            \end{scope}
            \begin{scope}
                \clip
                    (2.40,0.15)--(5.50,0.10)--(6.00,1.65)--
                    (5.15,3.40)--(3.25,3.00)--cycle;
                \fill[cyan!18]
                    (-0.6,-0.2)--(6.2,-0.2)--(6.2,1.16)--
                    (2.74,1.30)--(-0.6,1.43)--cycle;
            \end{scope}

            \draw[black!85,line width=0.9pt]
                (-0.15,0.25)--(2.40,0.15)--(3.25,3.00)--
                (1.45,3.35)--(-0.35,2.25)--cycle;
            \draw[black!85,line width=0.9pt]
                (2.40,0.15)--(5.50,0.10)--(6.00,1.65)--
                (5.15,3.40)--(3.25,3.00)--cycle;
            \draw[black,line width=1.25pt] (2.40,0.15)--(3.25,3.00);

            \draw[blue!65!black,line width=1.7pt]
                (-0.21,0.83)--(2.68,1.10)
                node[pos=0.30,above=-1pt] {$\Gamma_P^{\mathrm{rec}}$};
            \draw[blue!65!black,line width=1.7pt]
                (2.74,1.30)--(5.85,1.18)
                node[pos=0.78,above=-1pt] {$\Gamma_N^{\mathrm{rec}}$};
            \draw[cyan!55!black,line width=3.2pt]
                (2.40,0.15)--(2.68,1.10);

            \node[centre] at (1.30,1.72) {};
            \node[anchor=south east,inner sep=0pt,xshift=-1pt,yshift=1pt]
                at (1.30,1.72) {$\x_P$};
            \node[centre] at (4.45,1.72) {};
            \node[anchor=south east,inner sep=0pt,xshift=-1pt,yshift=1pt]
                at (4.45,1.72) {$\x_N$};
            \node at (0.96,2.70) {$\Omega_P$};
            \node at (5.15,2.75) {$\Omega_N$};
            \node[rotate=74,fill=white,inner sep=1pt] at (3.15,2.28)
                {$\mathcal S_f$};
            \node[rotate=74,fill=white,inner sep=0.8pt,
                text=cyan!45!black] at (2.53,0.66) {$\alpha_f^{(P)}$};
            \fill[cyan!18,draw=cyan!55!black] (0.15,-0.28) rectangle (0.42,-0.06);
            \node[anchor=west] at (0.50,-0.17) {liquid, $\alpha=1$};
            \fill[gray!8,draw=black!60] (3.65,-0.28) rectangle (3.92,-0.06);
            \node[anchor=west] at (4.00,-0.17) {gas, $\alpha=0$};
        \end{scope}

        \draw[gray!45] (6.75,-0.35)--(6.75,3.85);

        \begin{scope}[xshift=7.35cm]
            \node[font=\small\bfseries] at (3.0,3.85)
                {(b) Non-orthogonal decomposition};

            \path[cell]
                (0.0,0.30)--(2.55,0.18)--(3.35,3.12)--
                (1.50,3.38)--(-0.25,2.25)--cycle;
            \path[cell]
                (2.55,0.18)--(5.55,0.15)--(6.10,1.55)--
                (5.35,3.38)--(3.35,3.12)--cycle;
            \draw[black,line width=1.25pt] (2.55,0.18)--(3.35,3.12);

            \coordinate (P) at (1.30,1.76);
            \coordinate (N) at (5.00,1.68);
            \coordinate (F) at (2.95,1.65);
            \coordinate (D) at (4.50,1.59);
            \coordinate (S) at (4.50,1.02);

            \node[centre] at (P) {};
            \node[anchor=south east,inner sep=0pt,xshift=-1pt,yshift=1pt]
                at (P) {$\x_P$};
            \node[centre] at (N) {};
            \node[anchor=south west,inner sep=0pt,xshift=1pt,yshift=1pt]
                at (N) {$\x_N$};
            \node[centre,label=below left:$\x_f$] at (F) {};
            \node at (0.78,2.55) {$\Omega_P$};
            \node at (5.3,2.72) {$\Omega_N$};
            \node[rotate=75,fill=white,inner sep=1pt] at (3.22,2.45)
                {$\mathcal S_f$};

            \draw[->,black!70,line width=1.0pt]
                (P)--(N) node[pos=0.35,above=0pt] {$\mathbf d_f$};
            \draw[->,orange!85!black,dashed,line width=1.2pt]
                (F)--(D) node[pos=0.64,above=0pt] {$\boldsymbol{\Delta}_f$};
            \draw[->,green!45!black,line width=1.8pt]
                (F)--(S) node[pos=0.64,below=3pt] {$\Sf$};
            \draw[->,red!70!black,dash dot,line width=1.2pt]
                (D)--(S) node[midway,right] {$\mathbf k_f$};
            \node[font=\small] at (3.15,-0.20)
                {$\Sf=\boldsymbol{\Delta}_f+\mathbf k_f$};
        \end{scope}
    \end{tikzpicture}%
    }
    \caption{Geometric quantities used by the cell-centred unstructured FVM.
    (a) The cell-local planar reconstructions $\Gamma_P^{\mathrm{rec}}$ and
    $\Gamma_N^{\mathrm{rec}}$ partition the adjacent cells.  The highlighted
    liquid-covered part defines the candidate $\alpha_f^{(P)}$; the shared
    fraction combines the available cuts through \cref{eq:averagedFaceFraction}.
    (b) The owner and neighbour centroids are connected
    by $\mathbf d_f$, while
    $\Sf=\boldsymbol{\Delta}_f+\mathbf k_f$ separates the orthogonal and
    non-orthogonal parts of the face-area vector.  \schematicAidDisclosure}
    \label{fig:unstructuredFVMSchematic}
\end{figure}
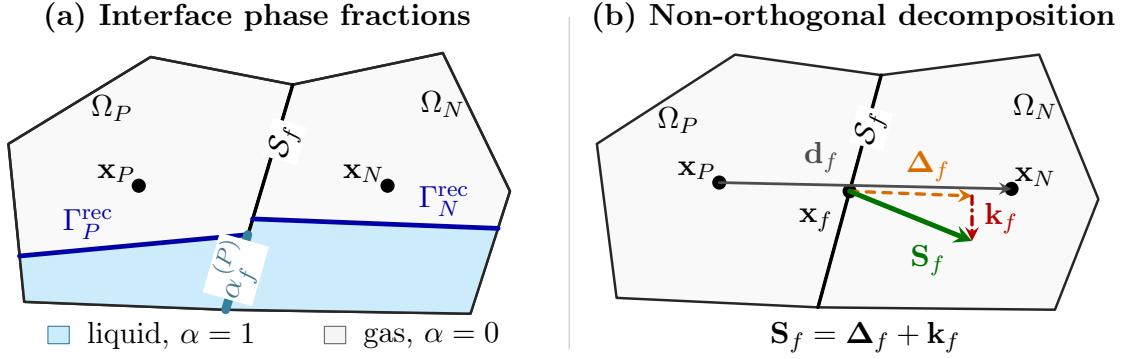

\subsection{Two-phase material representation}
\label{sec:materialRepresentation}

The two-phase material field is described by a volume fraction $\alpha$, with
$\alpha=1$ in the liquid, $\alpha=0$ in the gas, and intermediate values
in interface cells.  The mesh geometry, interface reconstruction, and
non-orthogonal decomposition are summarized in
\cref{fig:unstructuredFVMSchematic}.  The one-field density and acoustic
compressibility are linearly averaged as
\begin{equation}
    \rho=\alpha\rho_l+(1-\alpha)\rho_g,
    \qquad
    \betaac
    =\alpha\betaac_l+(1-\alpha)\betaac_g,
    \qquad
    \betaac_i=\frac{1}{\rho_i c_i^2},\quad i\in\{l,g\}.
    \label{eq:materialAverages}
\end{equation}

The cell fraction is the liquid volume fraction,
\begin{equation}
    \alpha_P=\frac{1}{V_P}\int_{\Omega_P}\chi_l\,dV,
    \label{eq:geometricFractions}
\end{equation}
where $\chi_l$ is the liquid indicator.  Face fractions are evaluated from the
available adjacent-cell piecewise-linear interface constructions (PLIC), using
the geometric cutting operations of the VOF reconstruction
\cite{Roenby2016,Scheufler2019,Liu2023incons}.
For an internal face $f$ shared by cells $P$ and $N$, let
$\mathcal C_f\subseteq\{P,N\}$ contain the cells with valid reconstructed
planes.  A plane with centre $\x_{\Gamma,c}$ and unit normal $\mathbf n_c$
defines the candidate liquid half-space
\[
    H_l^{(c)}
    =\{\x:\mathbf n_c\cdot(\x-\x_{\Gamma,c})\geq 0\}.
\]
The normal is oriented towards the liquid.  Each candidate supplies its own
liquid-covered area fraction,
\begin{equation}
    \alpha_f^{(c)}
    =\frac{\displaystyle\sum_{T\in\mathcal T_f}
                |T\cap H_l^{(c)}|}
           {\displaystyle\sum_{T\in\mathcal T_f}|T|},
    \qquad c\in\mathcal C_f ,
    \label{eq:candidateFaceFraction}
\end{equation}
where $\mathcal T_f$ is a triangulation of the face.  On a planar face this is
the liquid-covered area divided by the face area.  For a nonplanar polygon,
the denominator is the sum of actual triangle areas, rather than the
magnitude of the net oriented area vector.  The triangulation uses a canonical
vertex ordering and the first valid nonoverlapping vertex fan, so reversing
the face orientation on a processor boundary does not change the scalar
fraction.

The shared face fraction is
\begin{equation}
    \alpha_f=
    \begin{cases}
        \tfrac12\bigl(\alpha_f^{(P)}+\alpha_f^{(N)}\bigr),
            & |\mathcal C_f|=2,\\
        \alpha_f^{(c)}, & \mathcal C_f=\{c\}.
    \end{cases}
    \label{eq:averagedFaceFraction}
\end{equation}
Availability is determined by the reconstruction: a valid cut giving zero
or unit coverage remains a contribution.  When neither cell supplies a
plane, two pure cells of the same phase give the corresponding value zero
or one.  Opposite pure cells separated by a mesh face use the usual
distance-weighted linear interpolation of $\alpha_P$ and $\alpha_N$.
A cell classified as mixed by the reconstruction settings but lacking a
valid plane is treated as an error.  Nonfinite or inconsistent geometry is
rejected; only roundoff deviations are clamped to $[0,1]$.

On a processor face, the plane descriptions and their availability are
exchanged before the unsigned fractions are evaluated.  Both ranks form the
same average without using the sign of the volumetric flux $\phi$; their
agreement is checked to an absolute tolerance of $10^{-12}$.  Empty, wedge
and symmetry patches retain their boundary treatment.  Thus, face averaging
changes the material coefficient supplied to the acoustic operator without
introducing velocity-dependent donor selection or additional pressure
unknowns.

For the face flux, the same one-field relation is adopted as a closure for
the density on faces according to
\begin{equation}
    \rho_f=\alpha_f\rho_l+(1-\alpha_f)\rho_g,
    \qquad
    a_f:=\left(\frac{1}{\rho}\right)_f=\frac{1}{\rho_f}.
    \label{eq:faceInverseDensity}
\end{equation}
The PML tensors $\mathbf{T}_{0,f}$ and
$\mathbf{T}_{1,f}$ are linearly interpolated to faces; they are smooth in the
homogeneous PML and vanish in the heterogeneous physical domain.

\subsection{Interface flux}

The transmission conditions in \cref{eq:interfaceJumpConditions} identify
$a=1/\rho$ as the coefficient of the conserved normal pressure-gradient flux.
The submission method retains the one-field coefficient
$a_f=1/[\alpha_f\rho_l+(1-\alpha_f)\rho_g]$ from
\cref{eq:faceInverseDensity}.  For either pressure component $Q$, the
physical-domain face contribution is
\begin{equation}
    F_f(Q)=a_f\left[
        \frac{|\Sf|^2}{\Sf\cdot\mathbf d_f}(Q_N-Q_P)
        +(\nabla Q)_f\cdot\mathbf k_f
    \right],
    \label{eq:retainedInterfaceFlux}
\end{equation}
with the geometric decomposition and corrected gradient defined below.
The first term is implicit and the nonorthogonal correction is updated
explicitly.  A common face coefficient and gradient give opposite
contributions to the adjacent unscaled conservation equations, before
multiplication by their respective cell densities.

Geometric averaging does not replace this flux by separate liquid and gas
pressure-gradient reconstructions.  Consequently, the sharp-interface
conditions specify the continuum target, but are not imposed as exact local
constraints in a cut cell.  The discrete coefficient remains a one-field
closure whose accuracy must be assessed under interface displacement and
mesh refinement.  The liquid remains an acoustically transmitting medium;
normal acoustic velocity is continuous at the continuum interface, whereas
the complete phase-wise velocity vectors need not be identical.

\subsection{Control-volume balance}

For a scalar cell field $Q$, define the face flux associated with a tensor
coefficient $\mathbf{K}$ by
\begin{equation}
    \mathcal{F}_f(\mathbf{K},Q)
    :=\left(\mathbf{K}_f\cdot(\nabla Q)_f\right)\cdot\Sf.
    \label{eq:genericFaceFlux}
\end{equation}
Application of the Gauss theorem to \cref{eq:helmholtzPML} gives the two
cell balances
\begin{equation}
    (A_hP_{re})_P+(B_hP_{im})_P=0,
    \qquad
    (A_hP_{im})_P-(B_hP_{re})_P=0,
    \label{eq:discreteCellBalances}
\end{equation}
where
\begin{equation}
\begin{aligned}
    (A_hQ)_P={}&
        \rho_P\sum_{f\in\partial\Omega_P}
        a_f(\nabla Q)_f\cdot\Sf
        +\sum_{f\in\partial\Omega_P}
        \mathcal{F}_f(\mathbf{T}_0,Q)
        +(\omega^2\rho_P\betaac_P-C_{0,P})Q_PV_P,\\
    (B_hQ)_P={}&
        \sum_{f\in\partial\Omega_P}
        \mathcal{F}_f(\mathbf{T}_1,Q)
        +C_{1,P}Q_PV_P.
\end{aligned}
\label{eq:discreteOperators}
\end{equation}
The reaction coefficients and the orthogonal parts of the face fluxes are
assembled implicitly.

\subsection{Internal and boundary face gradients}

For an internal face shared by owner cell $P$ and neighbour cell $N$, let
$\mathbf{d}_f=\x_N-\x_P$.  The face-area vector is decomposed into an
orthogonal contribution $\boldsymbol{\Delta}_f$ and a non-orthogonal remainder
$\mathbf{k}_f$,
\begin{equation}
    \boldsymbol{\Delta}_f
    =\frac{|\Sf|^2}{\Sf\cdot\mathbf{d}_f}\mathbf{d}_f,
    \qquad
    \mathbf{k}_f=\Sf-\boldsymbol{\Delta}_f.
    \label{eq:faceAreaDecomposition}
\end{equation}
The corrected face-normal gradient is
\begin{equation}
    (\nabla Q)_f\cdot\Sf
    \simeq
    \frac{|\Sf|^2}{\Sf\cdot\mathbf{d}_f}(Q_N-Q_P)
    +(\nabla Q)_f\cdot\mathbf{k}_f.
    \label{eq:correctedFaceGradient}
\end{equation}
This decomposition follows the standard corrected unstructured FVM treatment
of non-orthogonal faces \cite{Jasak1996,Moukalled2016}.
Cell gradients are reconstructed with the selected unstructured gradient
scheme and interpolated linearly to the face.  The first term in
\cref{eq:correctedFaceGradient} contributes to the matrix coefficients.  When
a corrected Laplacian scheme is selected, the second term is evaluated
explicitly and updated in the non-orthogonal correction loop.  It vanishes on
an orthogonal mesh.  Tensor fluxes are evaluated from
$(\mathbf{T}_{m,f}\cdot\nabla Q_f)\cdot\Sf$ using the same face-gradient
reconstruction.

At a boundary face $b$, the flux is evaluated from the prescribed boundary
condition.  A Dirichlet condition supplies $Q_b$ and its face-normal gradient
is reconstructed from $Q_b-Q_P$, including the selected non-orthogonal
correction.  A Neumann condition directly prescribes
$g_b=\nabla Q_b\cdot\mathbf{n}_b$, giving the normal flux
$\rho_Pa_b|\mathcal{S}_b|g_b$.  In particular, a prescribed complex normal
velocity $V_n=V_{n,re}+iV_{n,im}$ gives, from the first-order momentum equation,
\begin{equation}
    \frac{\partial P_{re}}{\partial n}
    =-\omega\rho V_{n,im},
    \qquad
    \frac{\partial P_{im}}{\partial n}
    =\omega\rho V_{n,re}.
    \label{eq:velocityPressureBoundary}
\end{equation}
Here $V_n=\V\cdot\mathbf{n}$ is defined with the outward normal of the fluid
domain.  Consequently, a boundary moving into the fluid with real velocity
amplitude $u_0$ has $V_{n,re}=-u_0$ and is imposed through
$\partial P_{im}/\partial n=-\omega\rho u_0$.
For a Dirichlet condition, the coefficient multiplying the adjacent cell value
is added to the matrix diagonal, while the contribution containing the known
boundary value is transferred to the right-hand side.  Neumann data contribute
directly to the right-hand side.
Both contributions must be retained in the assembled block operator.

\subsection{Block system and correction sequence}

After assembly over all control volumes, the discrete equations have the
real-valued block form
\begin{equation}
\begin{bmatrix}
    A_h & B_h\\
    -B_h & A_h
\end{bmatrix}
\begin{bmatrix}
    \mathbf{P}_{re}\\
    \mathbf{P}_{im}
\end{bmatrix}
=
\begin{bmatrix}
    \mathbf{b}_{re}\\
    \mathbf{b}_{im}
\end{bmatrix}.
\label{eq:blockSystem}
\end{equation}
The discretization proceeds by reconstructing the interface and face fractions,
constructing the cell and face material coefficients, assembling the diagonal
and real--imaginary coupling operators, and solving \cref{eq:blockSystem}.
For a corrected non-orthogonal scheme, the explicit face-gradient correction
is refreshed and the block system is reassembled in each correction sweep.

For smooth coefficients and solutions on sufficiently regular meshes, linear
face interpolation and corrected gradients are expected to recover nominal
second-order spatial accuracy.  At a discontinuous material interface, the
accuracy also depends on the interface representation and the face coefficient
in \cref{eq:faceInverseDensity}.  Derived quantities based on reconstructed
pressure gradients can be more sensitive to mesh deformation than the primary
pressure solution.  These effects are examined in the corresponding
verification studies.

\section{Verification}
\label{sec:verification}

\subsection{Computational workflow and AI assistance}
\label{sec:aiWorkflow}

During this revision, OpenAI Codex (OpenAI, model
\texttt{gpt-6-astra}) assisted with source-code changes, preparation of
verification cases, data-processing and plotting scripts, and consistency
checks between the manuscript and its supporting data.  The corresponding
author directed these tasks and selected the changes included in the
revision.  Numerical results were obtained using the compiled solver or by
reprocessing archived simulation outputs, as distinguished in the code and
data availability statement.  The secondary-data package records the
inputs and scripts used to reproduce the numerical plots and tables.

\subsection{Error measures}
\label{sec:errorMeasures}

For analytical references, the numerical value $y_h(\x_i)$ is compared with
$y_{\mathrm{ex}}(\x_i)$ using
\begin{equation}
    E_2 =
    \frac{\left(\sum_i w_i|y_h(\x_i)-y_{\mathrm{ex}}(\x_i)|^2\right)^{1/2}}
    {\left(\sum_i w_i|y_{\mathrm{ex}}(\x_i)|^2\right)^{1/2}}.
    \label{eq:errorMeasures}
\end{equation}
Here $w_i=V_i$ for cell fields and $w_i=1$ for uniformly spaced sample
points.  With $\lVert q\rVert_w^2=\sum_iw_iq_i^2$ and
$e_q=q_h-q_{\mathrm{ex}}$, the component and combined complex-pressure errors
are computed as
\begin{equation}
    E_{P_{\mathrm{re}}}
    =\frac{\lVert e_{P_{\mathrm{re}}}\rVert_w}
           {\lVert P_{\mathrm{ex},\mathrm{re}}\rVert_w},
    \qquad
    E_{P_{\mathrm{im}}}
    =\frac{\lVert e_{P_{\mathrm{im}}}\rVert_w}
           {\lVert P_{\mathrm{ex},\mathrm{im}}\rVert_w},
    \qquad
    E_P
    =\left(
      \frac{\lVert e_{P_{\mathrm{re}}}\rVert_w^2
           +\lVert e_{P_{\mathrm{im}}}\rVert_w^2}
           {\lVert P_{\mathrm{ex},\mathrm{re}}\rVert_w^2
           +\lVert P_{\mathrm{ex},\mathrm{im}}\rVert_w^2}
      \right)^{1/2}.
    \label{eq:complexPressureErrors}
\end{equation}
Thus, $E_P$ combines the real and imaginary errors according to their
reference-field energies; it is not their arithmetic mean or maximum.
For two successive mesh sizes $h_1$ and $h_2$, the observed order is estimated
as
\begin{equation}
    q=\frac{\log(E_{h_1}/E_{h_2})}{\log(h_1/h_2)}.
    \label{eq:observedOrder}
\end{equation}
Six mesh levels are used for the homogeneous study, with the nominal mesh size
$h=\lambda/N_\lambda$.  The piston near and far fields are assessed on five
geometry-aligned meshes with 20--80 cells per
wavelength; the radial block boundary remains fixed at the piston edge.  For
the rigid-sphere position sweep, the force
difference is normalized by the peak
Gorkov force because a pointwise relative error is undefined at the analytical
zero-force positions.

\subsection{Homogeneous plane-wave baseline}
\label{sec:homogeneousBaseline}

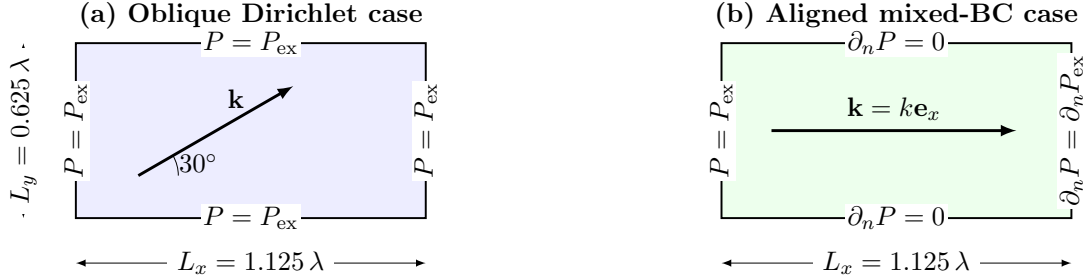
\begin{figure}[htbp]
    \centering
    \resizebox{0.96\textwidth}{!}{%
    \begin{tikzpicture}[x=0.72cm,y=0.72cm,>=latex,font=\small]
        \def\w{7.0}
        \def\h{3.5}

        \begin{scope}
            \fill[blue!7] (0,0) rectangle (\w,\h);
            \draw[thick] (0,0) rectangle (\w,\h);
            \node[font=\small\bfseries] at (3.5,4.05)
                {(a) Oblique Dirichlet case};
            \node[rotate=90,fill=white,inner sep=1pt] at (0,1.75)
                {$P=P_{\mathrm{ex}}$};
            \node[rotate=90,fill=white,inner sep=1pt] at (\w,1.75)
                {$P=P_{\mathrm{ex}}$};
            \node[fill=white,inner sep=1pt] at (3.5,0)
                {$P=P_{\mathrm{ex}}$};
            \node[fill=white,inner sep=1pt] at (3.5,\h)
                {$P=P_{\mathrm{ex}}$};
            \draw[->,very thick] (1.25,0.85) -- (4.35,2.64)
                node[pos=0.63,above] {$\mathbf{k}$};
            \draw (2.05,0.85) arc[start angle=0,end angle=30,radius=0.8];
            \node at (2.45,1.10) {$30^\circ$};
            \draw[<->] (0,-0.90) -- (\w,-0.90)
                node[midway,fill=white] {$L_x=1.125\, \lambda$};
            \draw[<->] (-1.05,0) -- (-1.05,\h)
                node[midway,fill=white,rotate=90] {$L_y=0.625\, \lambda$};
        \end{scope}

        \begin{scope}[xshift=9.3cm]
            \fill[green!7] (0,0) rectangle (\w,\h);
            \draw[thick] (0,0) rectangle (\w,\h);
            \node[font=\small\bfseries] at (3.5,4.05)
                {(b) Aligned mixed-BC case};
            \node[rotate=90,fill=white,inner sep=1pt] at (0,1.75)
                {$P=P_{\mathrm{ex}}$};
            \node[rotate=90,fill=white,inner sep=1pt] at (\w,1.75)
                {$\partial_n P=\partial_n P_{\mathrm{ex}}$};
            \node[fill=white,inner sep=1pt] at (3.5,0)
                {$\partial_nP=0$};
            \node[fill=white,inner sep=1pt] at (3.5,\h)
                {$\partial_nP=0$};
            \draw[->,very thick] (1.0,1.75) -- (5.9,1.75)
                node[midway,above] {$\mathbf{k}=k\mathbf{e}_x$};
            \draw[<->] (0,-0.90) -- (\w,-0.90)
                node[midway,fill=white] {$L_x=1.125\, \lambda$};
        \end{scope}

    \end{tikzpicture}
    }
    \caption{Boundary-condition configurations for the homogeneous plane-wave
    baseline: (a) oblique propagation with exact Dirichlet data and (b) aligned
    propagation with mixed Dirichlet--Neumann data.  Each configuration is
    solved on both mesh families shown in \cref{fig:homogeneousMeshes}.  \schematicAidDisclosure}
    \label{fig:homogeneousBaselineSchematic}
\end{figure}

The source-free baseline removes material and PML errors.  In a homogeneous
domain, the exact pressure is
\begin{equation}
    P(\x)=P_0\exp\!\left[i(\phi-\mathbf{k}\cdot\x)\right],
    \qquad |\mathbf{k}|=\omega/c_0.
    \label{eq:homogeneousPlaneWave}
\end{equation}
The domain dimensions are $L_x=1.125\, \lambda$ and
$L_y=0.625\, \lambda$, with $P_0=1\,\mathrm{Pa}$ and $\phi=\pi/6$.
The Dirichlet case uses a wave at $30^\circ$ to the $x$ axis and exact pressure
on all boundaries.  The mixed case uses an $x$-directed wave, exact pressure at
$x=0$, exact non-zero normal gradient at $x=L_x$, and zero normal gradient on
the lateral boundaries.  The exact velocity follows from
\cref{eq:velocityComplex}.  The two boundary configurations and mesh families
are summarized in \cref{fig:homogeneousBaselineSchematic}.

Two hexahedral mesh families are refined from 16 to 96 cells per wavelength.
The orthogonal family uses the orthogonal Laplacian.  In the non-orthogonal interior mesh
family, the boundary and first interior point lines remain Cartesian, giving
one orthogonal boundary-cell layer.  The deformation is ramped smoothly over
$0.15\min(L_x,L_y)$ to the prescribed interior deformation, whose amplitude
factor is $0.08$.  The resulting maximum non-orthogonality ranges from
$32.8^\circ$ to $37.2^\circ$, with average non-orthogonality between
$14.7^\circ$ and $16.0^\circ$.  The corrected Laplacian and two correction
sweeps are used.

\begin{figure}[htbp]
    \centering
    \begin{minipage}[t]{0.485\textwidth}
        \centering
        \includegraphics[width=\linewidth]{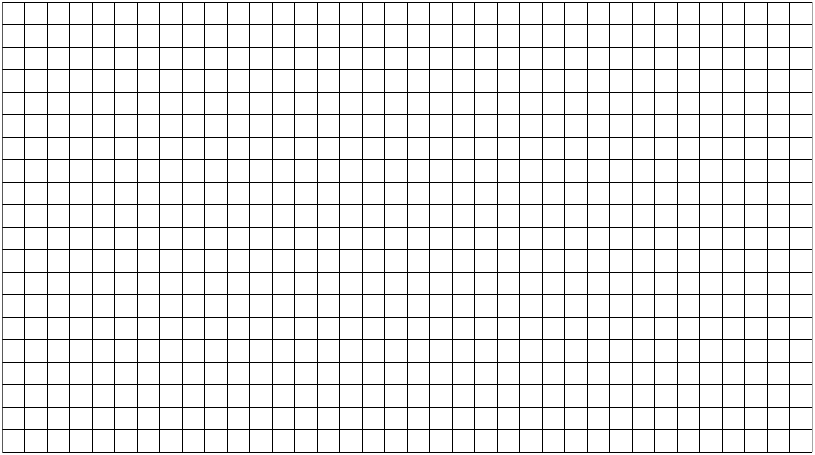}\\[-0.2em]
        \textbf{(a)} Orthogonal mesh
    \end{minipage}\hfill
    \begin{minipage}[t]{0.485\textwidth}
        \centering
        \includegraphics[width=\linewidth]{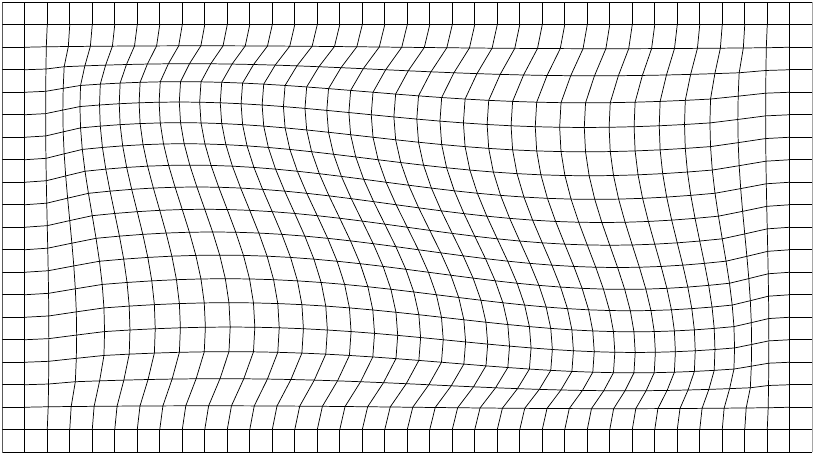}\\[-0.2em]
        \textbf{(b)} Non-orthogonal interior mesh
    \end{minipage}
    \caption{Representative homogeneous-baseline meshes at
    $N_\lambda=32$.  The non-orthogonal interior mesh retains one orthogonal
    boundary-cell layer and introduces the deformation smoothly towards the
    interior.  \schematicAidDisclosure}
    \label{fig:homogeneousMeshes}
\end{figure}

\begin{table}[htbp]
\centering
\small
\setlength{\tabcolsep}{4pt}
\begin{tabular}{llrrrrr}
\toprule
Mesh & BC & $N_\lambda$ & $E_2(P)$ & $q_P$ & $E_2(\mathbf{u})$ & $q_{\mathbf{u}}$ \\
\midrule
Orthogonal & Dirichlet & 32 & 4.421e-03 & 2.02 & 3.590e-03 & 1.98 \\
Orthogonal & Dirichlet & 48 & 1.958e-03 & 2.01 & 1.609e-03 & 1.98 \\
Orthogonal & Dirichlet & 64 & 1.100e-03 & 2.00 & 9.108e-04 & 1.98 \\
Orthogonal & Dirichlet & 96 & 4.885e-04 & 2.00 & 4.081e-04 & 1.98 \\
Orthogonal & Mixed & 32 & 1.301e-02 & 2.05 & 1.024e-02 & 2.04 \\
Orthogonal & Mixed & 48 & 5.722e-03 & 2.03 & 4.512e-03 & 2.02 \\
Orthogonal & Mixed & 64 & 3.207e-03 & 2.01 & 2.530e-03 & 2.01 \\
Orthogonal & Mixed & 96 & 1.421e-03 & 2.01 & 1.122e-03 & 2.01 \\
Interior-warped & Dirichlet & 32 & 1.026e-02 & 2.09 & 4.447e-02 & 1.50 \\
Interior-warped & Dirichlet & 48 & 4.662e-03 & 1.95 & 2.545e-02 & 1.38 \\
Interior-warped & Dirichlet & 64 & 2.564e-03 & 2.08 & 1.522e-02 & 1.79 \\
Interior-warped & Dirichlet & 96 & 1.121e-03 & 2.04 & 7.724e-03 & 1.67 \\
Interior-warped & Mixed & 32 & 1.843e-02 & 1.79 & 5.336e-02 & 1.51 \\
Interior-warped & Mixed & 48 & 9.606e-03 & 1.61 & 3.055e-02 & 1.38 \\
Interior-warped & Mixed & 64 & 5.449e-03 & 1.97 & 1.823e-02 & 1.79 \\
Interior-warped & Mixed & 96 & 2.481e-03 & 1.94 & 9.219e-03 & 1.68 \\
\bottomrule
\end{tabular}
\caption{Homogeneous plane-wave convergence. $N_\lambda$ denotes cells per wavelength; $q_P$ and $q_{\mathbf{u}}$ are the observed pressure and velocity orders.}
\label{tab:homogeneousConvergence}
\end{table}

\begin{figure}[htbp]
    \centering
    \includegraphics[width=0.95\textwidth]{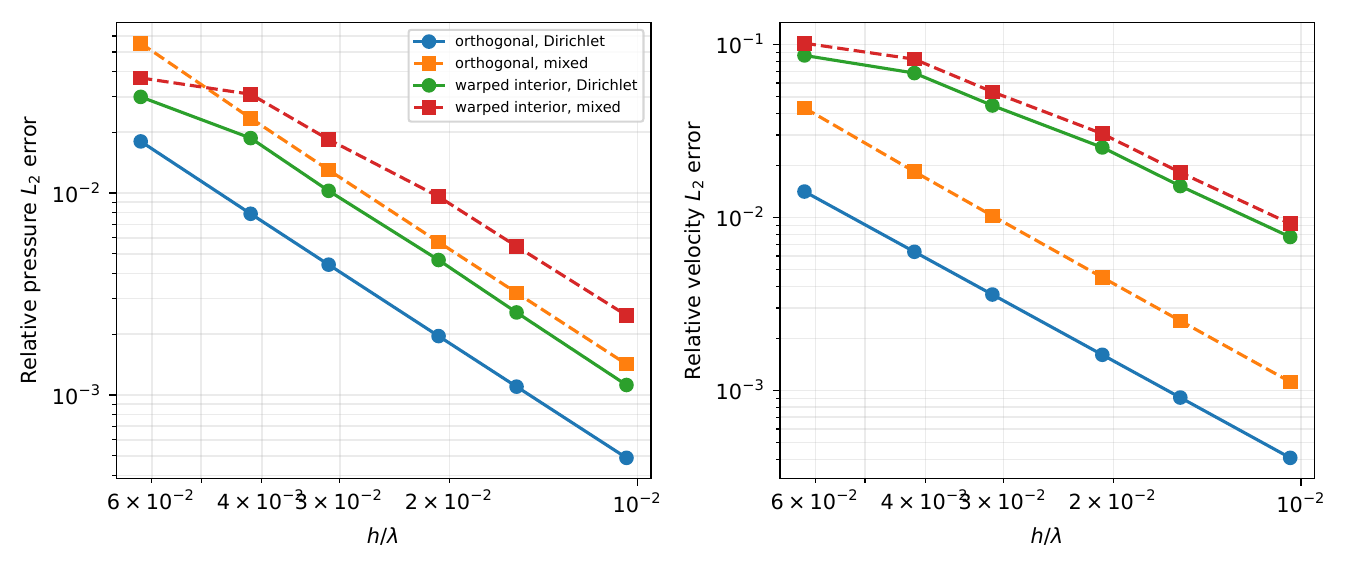}
    \caption{Homogeneous plane-wave pressure and velocity convergence on
    orthogonal and non-orthogonal interior meshes.  The latter retain one orthogonal
    boundary-cell layer.  \plotAidDisclosure}
    \label{fig:homogeneousConvergence}
\end{figure}

The orthogonal results give $q_P=2.00$--$2.01$ on the finest interval and
approximately second-order velocity convergence for both boundary treatments.
On the non-orthogonal interior mesh family, the finest pressure orders are $2.04$ for the
Dirichlet case and $1.94$ for the mixed case.  The corresponding velocity
orders are $1.67$ and $1.68$, showing that gradient reconstruction remains more
sensitive to mesh deformation than the primary pressure solution.  As a
diagnostic, deforming the boundary-adjacent cells as well as the interior
reduced the finest pressure orders to $1.03$ and $0.82$, respectively.  This
boundary-skewed variant is not used as the principal mesh family because it
mixes interior non-orthogonal flux errors with boundary-condition truncation
errors.

\subsection{Layered gas--liquid interface with PML}
\label{sec:layeredInterface}
\label{sec:combinedInterfacePML}

\begin{figure}[htbp]
    \centering
    \begin{tikzpicture}[x=1.15cm,y=1.1cm,>=latex,font=\small]
        \def\xzero{0}
        \def\xone{3.0}
        \def\xtwo{6.5}
        \def\xthree{9.0}
        \def\h{1.15}

        \fill[blue!10] (\xzero,0) rectangle (\xone,\h);
        \fill[green!12] (\xone,0) rectangle (\xtwo,\h);
        \fill[red!12] (\xtwo,0) rectangle (\xthree,\h);
        \draw[thick] (\xzero,0) rectangle (\xthree,\h);
        \draw[thick,dashed] (\xone,0) -- (\xone,\h);
        \draw[thick,dashed] (\xtwo,0) -- (\xtwo,\h);

        \draw[very thick] (\xzero,-0.08) -- (\xzero,\h+0.08);
        \draw[->,thick] (-0.85,0.58) -- (-0.25,0.58)
            node[midway,above] {$u_0e^{-i\omega t}$};

        \node at (1.5,0.62) {gas};
        \node at (1.5,0.25) {$\rho_1,\ c_1$};
        \node at (4.75,0.62) {liquid};
        \node at (4.75,0.25) {$\rho_2,\ c_2$};
        \node at (7.75,0.62) {liquid PML};
        \node at (7.75,0.25) {$\sigma_x(x)$};

        \draw[->,thick] (\xzero,-0.55) -- (9.55,-0.55) node[right] {$x$};
        \foreach \x/\lab in
            {\xzero/$0$,\xone/$L_1$,\xtwo/$L_2$,\xthree/$L$}
        {
            \draw (\x,-0.47) -- (\x,-0.65);
            \node[below] at (\x,-0.65) {\lab};
        }
        \node[above] at (\xone,\h) {interface};
        \node[above] at (\xtwo,\h) {PML start};
        \draw[black!70] (\xthree,0.58) -- (9.35,0.58);
        \node[anchor=west] at (9.42,0.58) {$\partial_nP=0$};
    \end{tikzpicture}
    \caption{Configuration of the integrated one-dimensional layered-interface
    and directional-PML validation case.  \schematicAidDisclosure}
    \label{fig:combinedLayeredSchematic}
\end{figure}
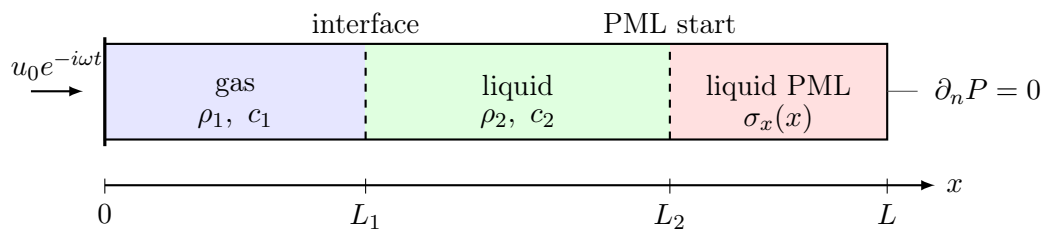

The heterogeneous material representation and the PML are assessed together
with a transducer-driven one-dimensional gas--liquid configuration.  A
prescribed normal velocity
$u_0=0.01\,\mathrm{m\,s^{-1}}$ at $x=0$ excites the gas at
$f=20\,\mathrm{kHz}$.  The gas--liquid interface is at
$L_1=0.09\,\mathrm{m}$, the liquid PML starts at $L_2=0.20\,\mathrm{m}$, and
the domain ends at $L=0.35\,\mathrm{m}$.  The gas properties are
$(\rho_1,c_1)=(1.2\,\mathrm{kg\,m^{-3}},343\,\mathrm{m\,s^{-1}})$ and the
liquid properties are
$(\rho_2,c_2)=(1000\,\mathrm{kg\,m^{-3}},1500\,\mathrm{m\,s^{-1}})$.  The PML uses
$\sigma_{\max}=5\times10^5\,\mathrm{s^{-1}}$ and polynomial order $p=3$.
The complete configuration is shown in \cref{fig:combinedLayeredSchematic}.

For the continuous directional PML, the analytical pressure is
\begin{equation}
P(x)=
\begin{cases}
\rho_1c_1u_0
\dfrac{e^{ik_1x}+Re^{ik_1(2L_1-x)}}{1-Re^{i2k_1L_1}},
&0<x<L_1,\\[2ex]
\rho_1c_1u_0
\dfrac{1+R}{1-Re^{i2k_1L_1}}
e^{i[k_2(x-L_1)+k_1L_1]},
&L_1<x<L_2,\\[2ex]
e^{-\frac{\sigma_{\max}(L-L_2)}{(p+1)c_2}
\left(\frac{x-L_2}{L-L_2}\right)^{p+1}}
\rho_1c_1u_0
\dfrac{1+R}{1-Re^{i2k_1L_1}}
e^{i[k_2(x-L_1)+k_1L_1]},
&L_2<x<L,
\end{cases}
\label{eq:combinedLayeredPMLAnalytical}
\end{equation}
where $k_j=\omega/c_j$ and
$R=(\rho_2c_2-\rho_1c_1)/(\rho_2c_2+\rho_1c_1)$.

\begin{figure}[htbp]
    \centering
    \includegraphics[width=0.78\textwidth]{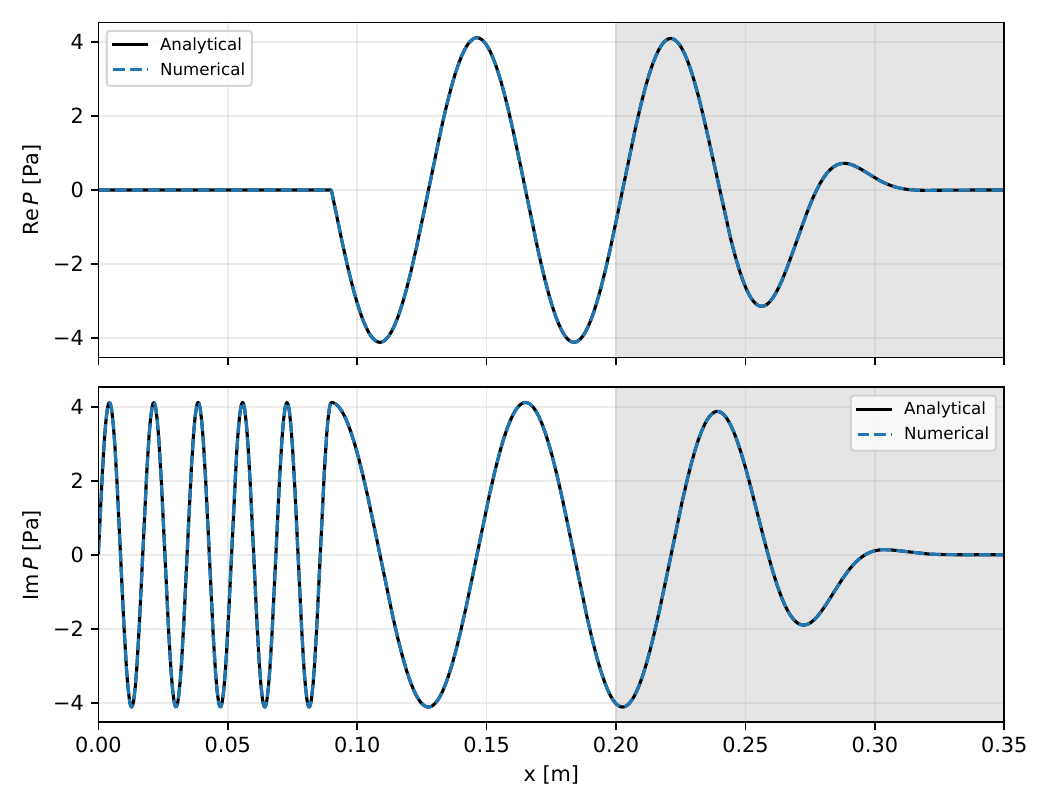}
    \caption{Integrated layered-interface--PML case: numerical real and
    imaginary pressure components compared with the analytical solution.  The
    shaded region denotes the PML.  \plotAidDisclosure}
    \label{fig:combinedLayeredPressure}
\end{figure}

\begin{figure}[htbp]
    \centering
    \includegraphics[width=0.78\textwidth]{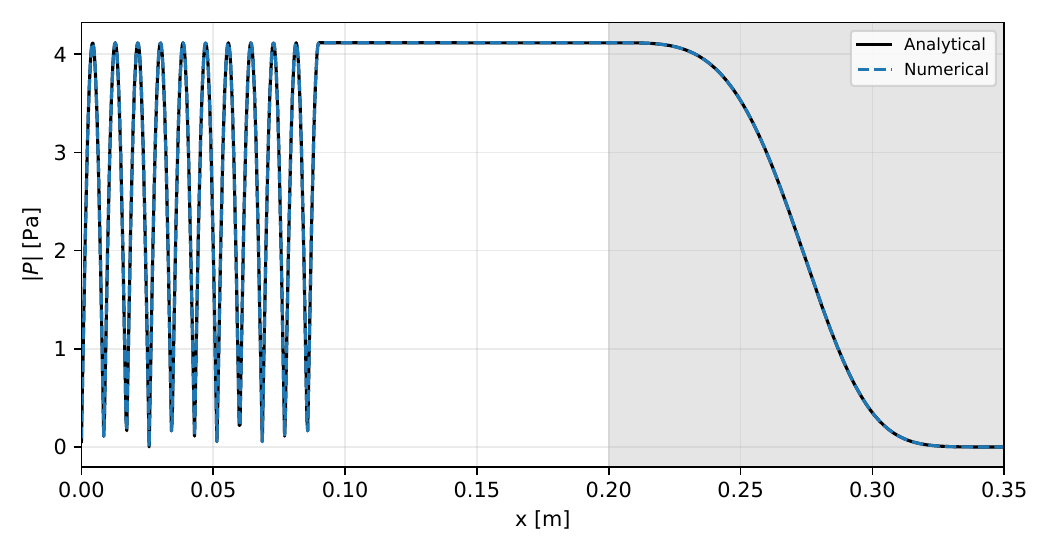}
    \caption{Integrated layered-interface--PML case: numerical and analytical
    pressure amplitude, including attenuation through the directional PML.  \plotAidDisclosure}
    \label{fig:combinedLayeredAmplitude}
\end{figure}

For the layered figures and sensitivity tables, the numerical cell fields are
converted to point data and linearly probed at 1200 uniformly spaced points
along the complete domain, including its endpoints.  The reported norm uses
unit sample weights and includes this interpolation; all samples must have
valid probe flags and finite pressure values.  It is distinct from a norm
computed directly at the cell centres.

On the $N_x=8000$ mesh, the complex-pressure error from
\cref{eq:errorMeasures}, evaluated over the complete interval $0\leq x\leq L$
including the PML, is $E_P=1.197\times10^{-3}$.  Here, $L_1$ lies inside an
interface cell rather than on a mesh face, so the comparison also exercises the
reconstructed face fraction in \cref{eq:faceInverseDensity}.  Thus, a single pointwise
comparison verifies the reflected and transmitted fields together with their
subsequent attenuation in the PML; no split into physical and absorbing
subdomains is used.  \cref{fig:combinedLayeredPressure} displays the real and
imaginary components needed to assess phase and amplitude simultaneously.
Although the complex comparison is sufficient quantitatively,
\cref{fig:combinedLayeredAmplitude} is retained because it makes the constant
transmitted amplitude and its smooth decay through the PML easier to interpret.

\begin{figure}[H]
    \centering
    \includegraphics[width=0.70\textwidth]{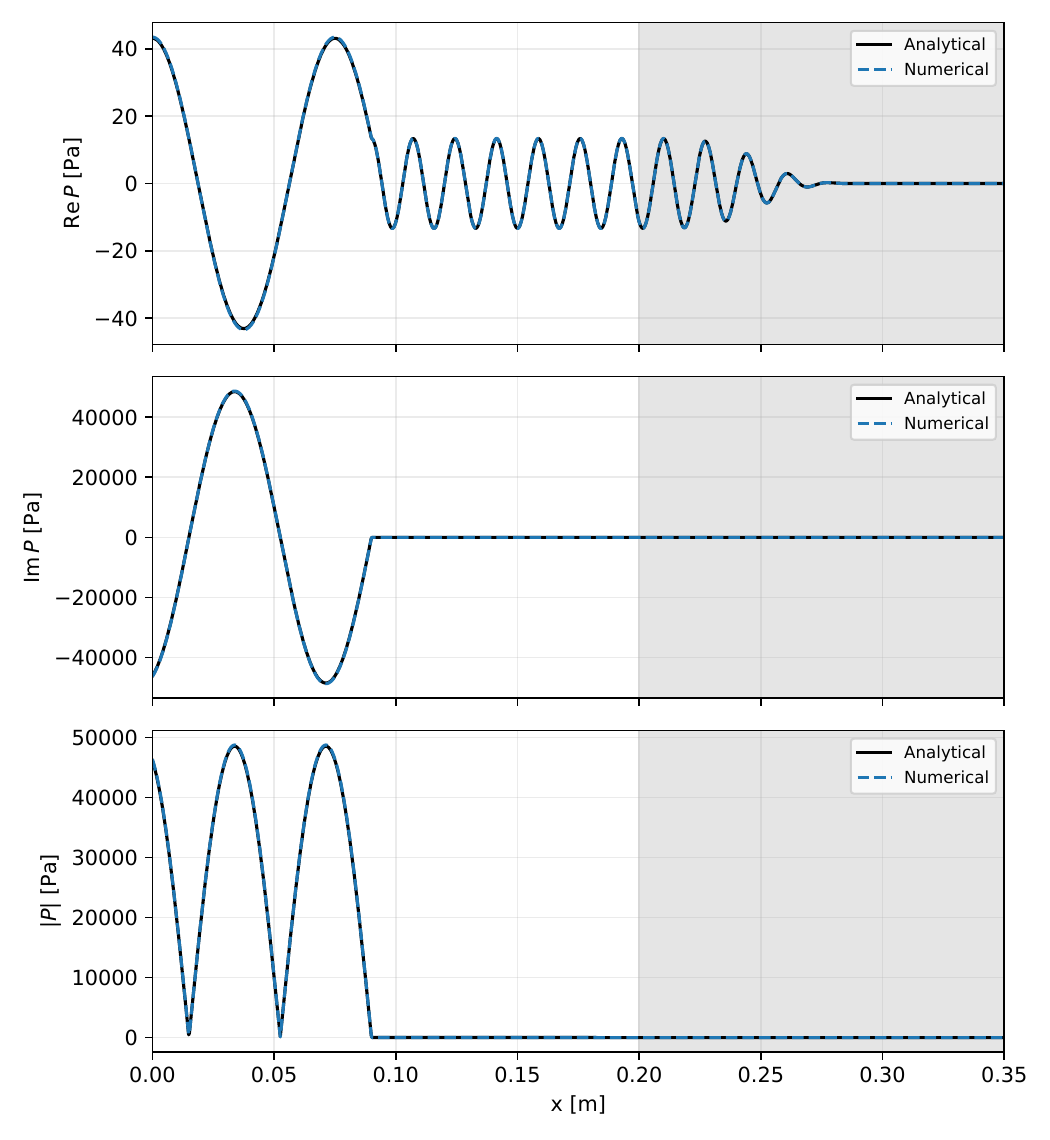}
    \caption{Reverse layered-interface--PML check: numerical and analytical
    pressure for propagation from water into air.  The upper panels show the
    real and imaginary components, and the lower panel shows the amplitude.
    The pressure approaches a node at the interface; the shaded region denotes
    the air-side PML.  \plotAidDisclosure}
    \label{fig:combinedLayeredWaterAir}
\end{figure}

As an additional reverse-incidence check, the two materials are interchanged so
that the wave travels from water into air, while the geometry, excitation, and
PML parameters remain unchanged.  The pressure reflection coefficient is then
$R=-0.999451$, corresponding to an almost complete phase-reversed reflection
and a much smaller transmitted pressure in air.  The numerical and analytical
complex pressures remain in close agreement over the complete domain in
\cref{fig:combinedLayeredWaterAir}, with $E_P=4.425\times10^{-3}$.

\FloatBarrier
\subsubsection{PML damping and mesh sensitivity}
\label{sec:pmlSensitivity}

\begin{figure}[htbp]
    \centering
    \includegraphics[width=0.75\textwidth]{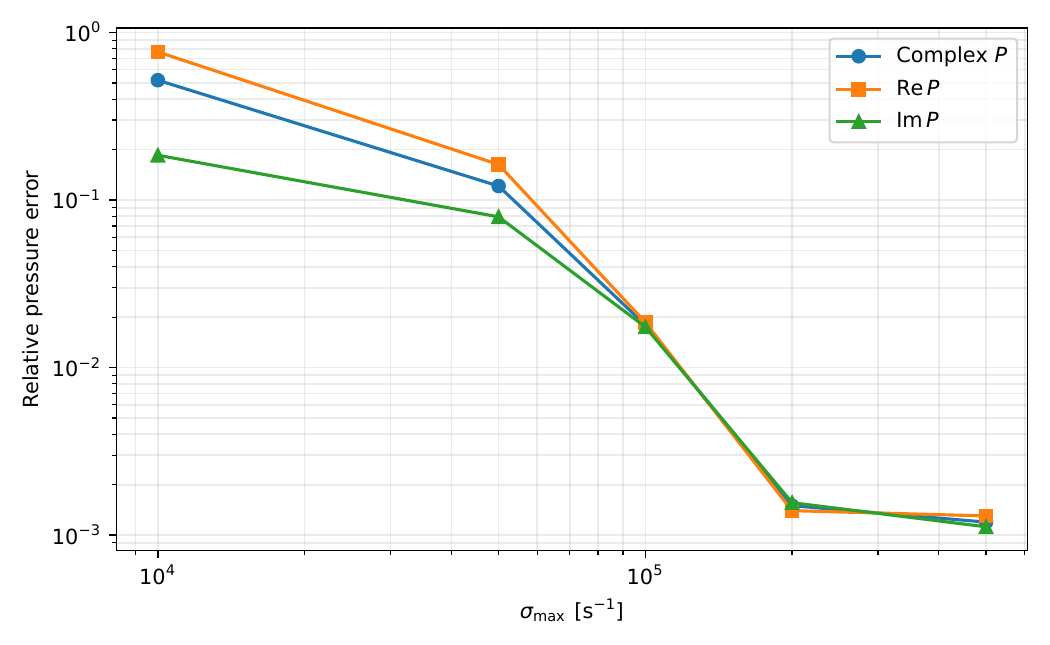}
    \caption{Whole-domain complex-pressure and component errors for the layered
    gas--liquid--PML case as functions of maximum PML damping.  \plotAidDisclosure}
    \label{fig:pmlSigmaSensitivity}
\end{figure}

The PML parameters are studied with the same layered configuration as in \cref{fig:combinedLayeredSchematic} and the
same whole-domain complex-pressure error.  First, $N_x=8000$, the PML thickness,
and $p=3$ are held fixed while $\sigma_{\max}$ is varied.  The results in
\cref{tab:pmlSigmaSensitivity,fig:pmlSigmaSensitivity} show that weak damping
does not sufficiently attenuate the wave before the outer boundary.  Increasing
$\sigma_{\max}$ from $10^4$ to $5\times10^5\,\mathrm{s^{-1}}$ reduces $E_P$
from $5.188\times10^{-1}$ to $1.197\times10^{-3}$.  The component errors are
shown alongside $E_P$ to expose differences in the real and imaginary fields;
at the selected damping they are $1.303\times10^{-3}$ and
$1.122\times10^{-3}$, respectively.

\begin{table}[htbp]
    \centering
    \caption{Directional tensor PML damping sensitivity for the one-dimensional layered-interface case.}
    \label{tab:pmlSigmaSensitivity}
    \begin{tabular}{rrrr}
        \toprule
        $\sigma_{\max}$ [$\mathrm{s}^{-1}$] & $E_P$ & $E_{P_{\mathrm{re}}}$ & $E_{P_{\mathrm{im}}}$ \\
        \midrule
        10000 & $5.188\times10^{-1}$ & $7.655\times10^{-1}$ & $1.851\times10^{-1}$ \\
        50000 & $1.212\times10^{-1}$ & $1.629\times10^{-1}$ & $7.937\times10^{-2}$ \\
        100000 & $1.797\times10^{-2}$ & $1.865\times10^{-2}$ & $1.750\times10^{-2}$ \\
        200000 & $1.500\times10^{-3}$ & $1.401\times10^{-3}$ & $1.563\times10^{-3}$ \\
        500000 & $1.197\times10^{-3}$ & $1.303\times10^{-3}$ & $1.122\times10^{-3}$ \\
        \bottomrule
    \end{tabular}
\end{table}

\begin{figure}[htbp]
    \centering
    \includegraphics[width=0.75\textwidth]{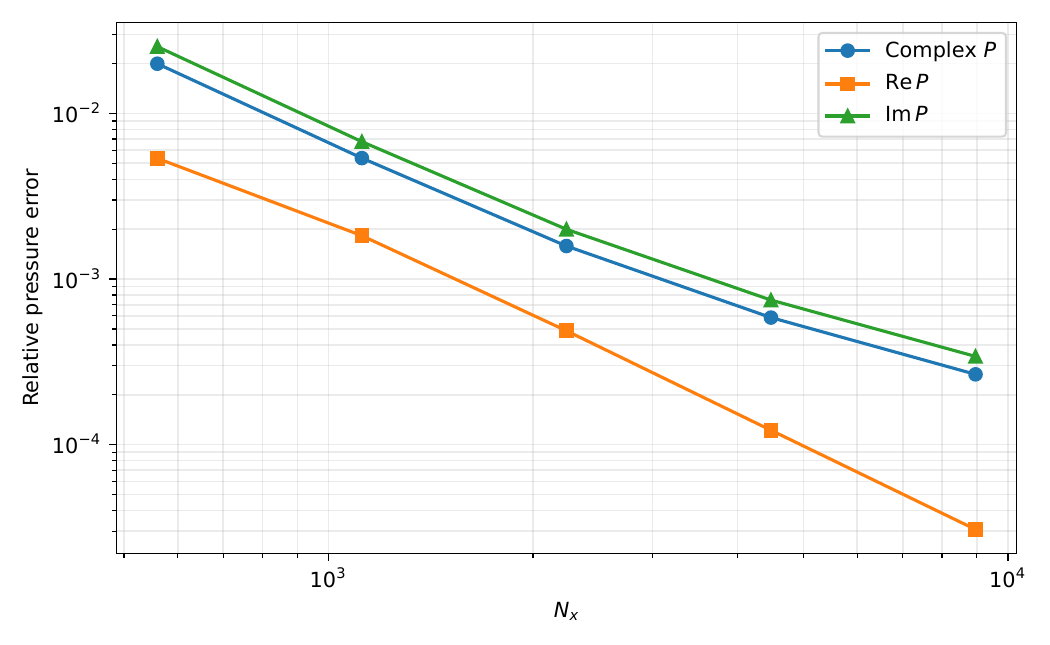}
    \caption{Whole-domain complex-pressure and component errors under aligned
    mesh refinement for the layered gas--liquid--PML case.  \plotAidDisclosure}
    \label{fig:pmlMeshSensitivity}
\end{figure}

For the resolution study, $\sigma_{\max}=5\times10^5\,\mathrm{s^{-1}}$ is
fixed and the mesh is refined while keeping the material interface and PML
start aligned with mesh faces.  \Cref{tab:pmlMeshSensitivity,fig:pmlMeshSensitivity}
show a monotonic decrease of $E_P$ from $1.997\times10^{-2}$ at $N_x=560$ to
$2.657\times10^{-4}$ at $N_x=8960$, demonstrating convergence of the complete
coupled solution.  Both component errors also decrease monotonically over the
same sequence.

\begin{table}[H]
    \centering
    \caption{Directional tensor PML mesh sensitivity for the one-dimensional layered-interface case with fixed damping $\sigma_{\max}=5.0\times10^5\,\mathrm{s^{-1}}$.}
    \label{tab:pmlMeshSensitivity}
    \begin{tabular}{rrrrr}
        \toprule
        $N_x$ & $\Delta x$ [$\mathrm{m}$] & $E_P$ & $E_{P_{\mathrm{re}}}$ & $E_{P_{\mathrm{im}}}$ \\
        \midrule
        560 & $6.250\times10^{-4}$ & $1.997\times10^{-2}$ & $5.360\times10^{-3}$ & $2.533\times10^{-2}$ \\
        1120 & $3.125\times10^{-4}$ & $5.376\times10^{-3}$ & $1.827\times10^{-3}$ & $6.757\times10^{-3}$ \\
        2240 & $1.563\times10^{-4}$ & $1.582\times10^{-3}$ & $4.870\times10^{-4}$ & $1.997\times10^{-3}$ \\
        4480 & $7.813\times10^{-5}$ & $5.853\times10^{-4}$ & $1.221\times10^{-4}$ & $7.465\times10^{-4}$ \\
        8960 & $3.906\times10^{-5}$ & $2.657\times10^{-4}$ & $3.074\times10^{-5}$ & $3.409\times10^{-4}$ \\
        \bottomrule
    \end{tabular}
\end{table}

\FloatBarrier
\subsection{Controlled comparison of face-area definitions}
\label{sec:geometricComparison}

To isolate the effect of \cref{eq:averagedFaceFraction}, we compare the
original donor-based face-area selection with geometric averaging while
retaining the flux in \cref{eq:retainedInterfaceFlux}.  Cell-density
interpolation, reaction quadrature, boundary conditions, PML coefficients,
and velocity postprocessing are identical between the two selections.
Each pair uses the same prepared mesh and initial interface.

The layered cases in \cref{sec:layeredInterface,sec:pmlSensitivity} were
repeated with both selections: both material orderings at $N_x=8000$, all
five damping values, and all five aligned mesh sizes.  The serial complex
pressure fields agree to roundoff, and the largest relative difference
from the original serial field in the serial and parallel checks is
$1.95\times10^{-12}$.  These checks comprise 44 runs on one/four ranks
and 32 runs on one/eight ranks; the three intermediate damping values
were checked on one/four ranks.  The same 1200-point sampling is applied
to the selected-method fields for all layered figures and sensitivity tables.

A curved-interface comparison uses a penetrable liquid sphere of radius
$a=1\,\mathrm{mm}$ centred in the cube $[-2a,2a]^3$.  The liquid and gas
properties are those of \cref{sec:layeredInterface}.  An incident plane wave
$P_{\mathrm{inc}}=1\,\mathrm{Pa}\,e^{ik_gx}$ has $k_ga=1$.
The independent reference matches regular interior spherical-Bessel and
outgoing exterior spherical-Hankel expansions through continuity of
pressure and of $\rho^{-1}\partial_nP$ at $r=a$, retaining angular orders
$0$--$24$.  Its total complex pressure supplies Dirichlet data on the
cube boundary, so no PML is used in this comparison.  The cell-centred
numerical pressure is compared with this reference using cell-volume
weights.  Decomposition dependence is measured after reconstructing the
parallel field in the original cell ordering as
\begin{equation}
 D_8=\frac{\lVert P_h^{(8)}-P_h^{(1)}\rVert_V}
                {\lVert P_h^{(1)}\rVert_V}.
 \label{eq:parallelPressureDifference}
\end{equation}

\begin{table}[htbp]
\centering
\small
\setlength{\tabcolsep}{5pt}
\begin{tabular}{rcccc}
\toprule
& \multicolumn{2}{c}{$E_P$ (serial)} & \multicolumn{2}{c}{$D_8$} \\
$N$ & Original & Geometric & Original & Geometric \\
\midrule
16 & $9.2118\times10^{-1}$ & $9.1199\times10^{-1}$ & $1.8520\times10^{-14}$ & $1.5313\times10^{-14}$ \\
24 & $1.7960\times10^{-1}$ & $1.7983\times10^{-1}$ & $2.9721\times10^{-4}$ & $1.4119\times10^{-15}$ \\
32 & $6.4716\times10^{-2}$ & $6.4767\times10^{-2}$ & $3.1739\times10^{-5}$ & $1.8214\times10^{-15}$ \\
48 & $2.9746\times10^{-2}$ & $2.9753\times10^{-2}$ & $2.6395\times10^{-6}$ & $4.9163\times10^{-15}$ \\
\bottomrule
\end{tabular}
\caption{Controlled comparison of the original face-area selection and
geometric averaging for a penetrable liquid sphere on $N^3$ Cartesian cells.
Both methods use the same pressure flux. $E_P$ is the volume-weighted
complex-pressure error against the series solution; $D_8$ compares eight-rank
and serial pressure fields for the same method.}
\label{tab:geometricComparison}
\end{table}

\Cref{tab:geometricComparison} reports all four tested resolutions.
On the finest mesh, geometric averaging changes $E_P$ from $2.9746\%$
to $2.9753\%$, while reducing $D_8$ from $2.64\times10^{-6}$ to
$4.92\times10^{-15}$.  Both selections retain 3,041,280 matrix entries.
The true relative assembled residual is below $10^{-10}$ in these
sphere and selected-method layered runs.  These results motivate the
selection through reproducibility with essentially unchanged pressure
accuracy; they do not establish improved force accuracy or a general
convergence order for curved interfaces.

\FloatBarrier
\subsection{Baffled-piston radiation}
\label{sec:piston}

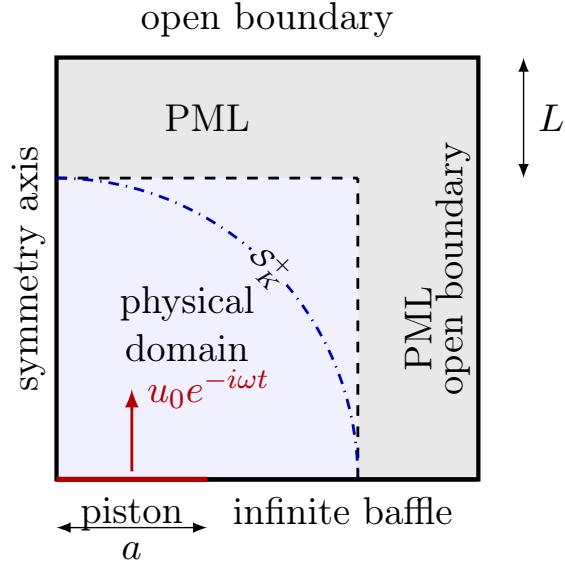
\begin{figure}[H]
    \centering
    \resizebox{0.52\textwidth}{!}{%
    \begin{tikzpicture}[x=1cm,y=1cm,>=latex,font=\small]
        \def\inner{3.0}
        \def\outer{4.2}
        \def\piston{1.5}

        \fill[blue!6] (0,0) rectangle (\inner,\inner);
        \fill[gray!18] (\inner,0) rectangle (\outer,\outer);
        \fill[gray!18] (0,\inner) rectangle (\inner,\outer);
        \draw[very thick] (0,0) rectangle (\outer,\outer);
        \draw[dashed,thick] (\inner,0) -- (\inner,\inner);
        \draw[dashed,thick] (0,\inner) -- (\inner,\inner);
        \draw[dash dot,blue!70!black,thick]
            (0,\inner) arc[start angle=90,end angle=0,radius=\inner];

        \draw[red!70!black,line width=1.5pt] (0,0) -- (\piston,0);
        \draw[very thick] (\piston,0) -- (\outer,0);
        \draw[<->] (0,-0.48) -- (\piston,-0.48)
            node[midway,below] {$a$};
        \draw[->,red!70!black,thick] (0.75,0.08) -- (0.75,0.90)
            node[right] {$u_0e^{-i\omega t}$};

        \draw[<->] (4.65,\inner) -- (4.65,\outer)
            node[midway,right] {$L$};

        \node[rotate=90] at (-0.28,2.1) {symmetry axis};
        \node[align=center] at (1.3,1.55) {physical\\domain};
        \node[rotate=-45,fill=blue!6,inner sep=1pt,font=\scriptsize]
            at (2.12,2.12) {$\mathcal{S}_K^+$};
        \node[rotate=90] at (3.60,1.50) {PML};
        \node at (1.50,3.60) {PML};
        \node[below] at (0.75,-0.03) {piston};
        \node[below] at (2.85,-0.03) {infinite baffle};
        \node[above] at (2.1,{\outer+0.06}) {open boundary};
        \node[rotate=90,above] at ({\outer+0.06},2.1) {open boundary};
    \end{tikzpicture}
    }
    \caption{Axisymmetric baffled-piston configuration.  The rectangular PML
    occupies the outer radial and upper strips and their corner.  The
    dash-dotted arc is the upper Kirchhoff reconstruction auxiliary surface and not belong to any boundary.  \schematicAidDisclosure}
    \label{fig:pistonConfiguration}
\end{figure}

The baffled-piston radiation problem provides analytical references for the
on-axis near field and far-field pressure pattern \cite{oelze2003piston}.  A rigid
circular piston of radius $a$ oscillates with velocity amplitude $u_0$ in an
infinite baffle.  The numerical domain is truncated by a rectangular PML.
The case uses $f=10\,\mathrm{kHz}$, $a=0.1\,\mathrm{m}$,
$(\rho_0,c)=(1.2\,\mathrm{kg\,m^{-3}},343\,\mathrm{m\,s^{-1}})$, a
$2^\circ$ axisymmetric wedge, and a rectangular meridional domain
$0\leq x,y\leq0.28\,\mathrm{m}$.  The PML starts at $x=0.20\,\mathrm{m}$ and
$y=0.20\,\mathrm{m}$, giving a thickness of $0.08\,\mathrm{m}$ along the outer
radial and upper boundaries.  To preserve the geometry during refinement, a
radial block boundary is placed exactly at $x=a$ and directly separates the
driven piston patch from the rigid baffle.  The geometry is shown in
\cref{fig:pistonConfiguration}.

The on-axis near-field pressure magnitude is
\begin{equation}
    |p|(r,\theta=0)
    =
    2\rho_0cu_0
    \left|
    \sin\left\{
    \frac{1}{2}kr\left[\sqrt{1+\left(\frac{a}{r}\right)^2}-1\right]
    \right\}
    \right|.
    \label{eq:pistonNearField}
\end{equation}
In the far field, the analytical complex pressure amplitude, written with the
$e^{-i\omega t}$ convention of \cref{as:timeHarmonic}, is
\begin{equation}
    P(r,\theta)
    =
    -\frac{i\omega a^2\rho_0u_0}{2r}e^{ikr}
    \left[
    \frac{2J_1(ka\sin\theta)}{ka\sin\theta}
    \right].
    \label{eq:pistonFarField}
\end{equation}

The far field is reconstructed from the numerical solution rather than meshed
directly.  Let $\mathcal{S}_K^+$ denote the upper hemispherical reconstruction
surface of radius $r_K=0.20\,\mathrm{m}$.
The infinite rigid baffle is included by sound-hard symmetry.  With
$\mathcal{R}_b$ denoting reflection across the baffle plane, the complete
source surface is
\begin{equation}
    \mathcal{S}_K
    =
    \mathcal{S}_K^+
    \cup\mathcal{R}_b(\mathcal{S}_K^+),
    \qquad
    P(\mathcal{R}_b\mathbf{y})=P(\mathbf{y}),
    \qquad
    \frac{\partial P}{\partial n}
    (\mathcal{R}_b\mathbf{y})
    =
    \frac{\partial P}{\partial n}(\mathbf{y}).
    \label{eq:soundHardMirror}
\end{equation}
Thus, $\mathcal{S}_K$ is closed and encloses the piston source.  Let
$\mathbf{n}_{\mathbf{y}}$ point from the enclosed source region towards the
exterior.  The implemented full, finite-distance Kirchhoff--Helmholtz
reconstruction is
\begin{equation}
    P_K(\mathbf{x})
    =\int_{\mathcal{S}_K}
    \left[
        P(\mathbf{y})
        \frac{\partial G(\mathbf{x},\mathbf{y})}
             {\partial n_{\mathbf{y}}}
        -G(\mathbf{x},\mathbf{y})
        \frac{\partial P(\mathbf{y})}{\partial n_{\mathbf{y}}}
    \right]dS_{\mathbf{y}},
    \label{eq:kirchhoffReconstruction}
\end{equation}
where, for the $e^{-i\omega t}$ convention,
\begin{equation}
    \begin{aligned}
        G(\mathbf{x},\mathbf{y})
        &=\frac{e^{ikR}}{4\pi R},
        &
        \frac{\partial G}{\partial n_{\mathbf{y}}}
        &=\left(\frac{1}{R}-ik\right)G
          \widehat{\mathbf{R}}\cdot\mathbf{n}_{\mathbf{y}},\\
        \mathbf{R}&=\mathbf{x}-\mathbf{y},
        &R&=|\mathbf{R}|,
        \qquad
        \widehat{\mathbf{R}}=\frac{\mathbf{R}}{R}.
    \end{aligned}
    \label{eq:kirchhoffGreenFunction}
\end{equation}
This representation follows the standard exterior Helmholtz boundary integral
\cite{morse1986theoretical}.

The numerical complex pressure and its consistently reconstructed gradient are
interpolated to polar samples $\mathbf{y}_j$ on the upper meridional contour.
The samples are reflected according to \cref{eq:soundHardMirror}, giving
$0\leq\vartheta_j\leq\pi$.  Revolving the completed contour through
$0\leq\varphi_l<2\pi$ gives the three-dimensional quadrature
\begin{equation}
    P_K(\mathbf{x}_m)
    \approx
    \sum_j\sum_l
    \left[
        P_{jl}\left(\frac{\partial G}{\partial n}\right)_{mjl}
        -G_{mjl}(\nabla P)_{jl}\cdot\mathbf{n}_{jl}
    \right]
    r_K^2\sin\vartheta_j\,\Delta\vartheta_j\Delta\varphi,
    \label{eq:kirchhoffQuadrature}
\end{equation}
with $\nabla P=\nabla P_{\mathrm{re}}+i\nabla P_{\mathrm{im}}$.  The evaluation
points $x_m$ are placed at $R=4\,\mathrm{m}$ and $0^\circ\leq\theta\leq90^\circ$.
For visualization, the reconstructed and analytical pressures are plotted as
$L_p=20\log_{10}(|P|/p_{\mathrm{ref}})$, with
$p_{\mathrm{ref}}=20\,\mu\mathrm{Pa}$.  Over the uniformly spaced observation
angles, the far-field pressure-magnitude error is
\begin{equation}
    E_{2,|P|}^{\mathrm{ff}}
    =\frac{\left[\sum_m
        \left(|P_K(R,\theta_m)|-|P_{\mathrm{ex}}(R,\theta_m)|\right)^2
        \right]^{1/2}}
        {\left[\sum_m|P_{\mathrm{ex}}(R,\theta_m)|^2\right]^{1/2}}.
    \label{eq:pistonFarFieldErrors}
\end{equation}
This measure retains the pressure-amplitude error and is used to assess mesh
refinement.

\begin{table}[htbp]
\centering
\begin{tabular}{cccc}
\hline
Cells per wavelength & $h/\lambda$ & $E_2$ & $E_\infty$ \\
\hline
20 & 0.0500 & 2.594e-02 & 4.204e-02 \\
30 & 0.0333 & 1.208e-02 & 3.604e-02 \\
40 & 0.0250 & 7.197e-03 & 3.038e-02 \\
60 & 0.0167 & 3.737e-03 & 2.261e-02 \\
80 & 0.0125 & 2.579e-03 & 1.802e-02 \\
\hline
\end{tabular}
\caption{Mesh-convergence study for the piston-radiation near-field pressure amplitude along the symmetry axis.}
\label{tab:pistonConvergence}
\end{table}

\begin{figure}[H]
    \centering
    \includegraphics[width=0.85\textwidth]{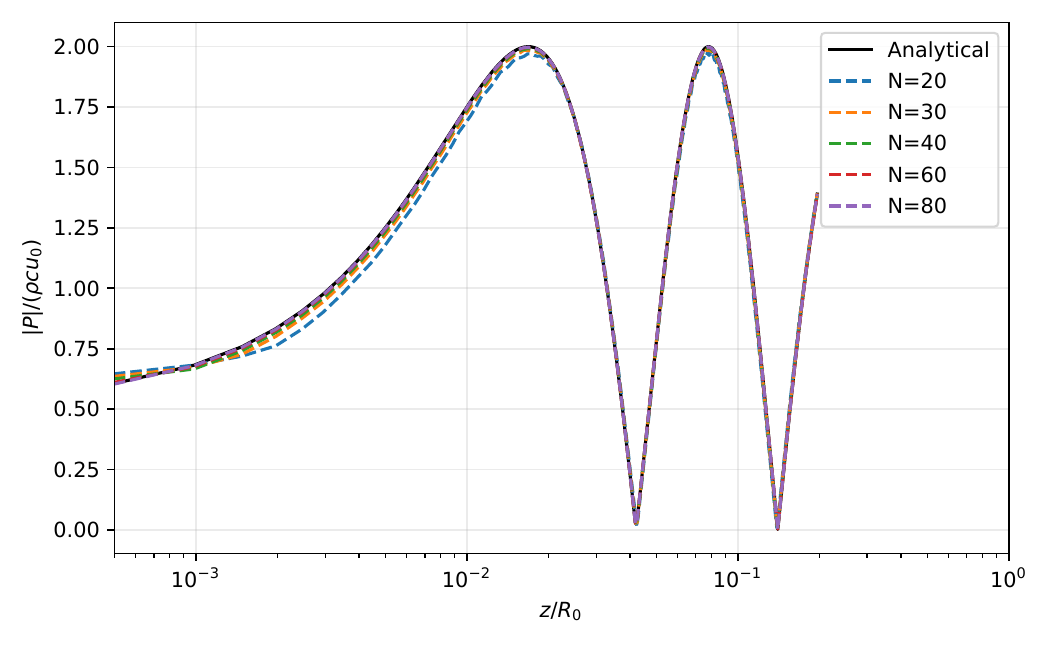}
    \caption{Baffled-piston radiation: on-axis near-field pressure amplitude
    compared with the analytical solution.  \plotAidDisclosure}
    \label{fig:pistonNearAmpMethod}
\end{figure}

\begin{figure}[H]
    \centering
    \includegraphics[width=0.85\textwidth]{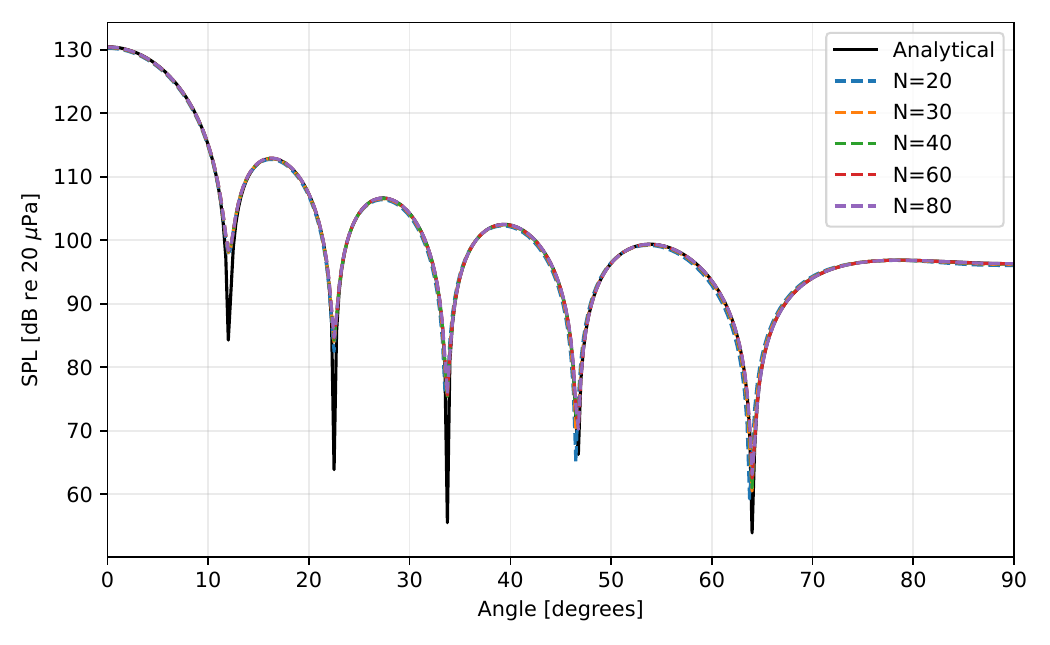}
    \caption{Baffled-piston radiation: far-field sound pressure level
    reconstructed on five mesh densities and compared with the analytical
    pressure pattern.  \plotAidDisclosure}
    \label{fig:pistonFarAmpMethod}
\end{figure}

For the finest mesh, the far-field directivity pattern is additionally shown
in polar form.  The analytical pressure directivity follows from
\cref{eq:pistonFarField} as
\begin{equation}
    D(\theta)
    =\frac{P(R,\theta)}{P(R,0)}
    =\frac{2J_1(ka\sin\theta)}{ka\sin\theta},
    \qquad D(0)=1,
    \label{eq:pistonDirectivity}
\end{equation}
and the plotted directivity level is
\begin{equation}
    D_{\mathrm{dB}}(\theta)
    =20\log_{10}\left|
      \frac{P(R,\theta)}{P(R,0)}
     \right|
    =20\log_{10}|D(\theta)|.
    \label{eq:pistonDirectivityLevel}
\end{equation}
The pressure ratio is evaluated from the reconstructed numerical field, while
the final expression applies to the analytical solution.  This representation
makes the main lobe, side lobes, and pressure minima directly visible without
introducing a second error measure.

\begin{figure}[H]
    \centering
    \includegraphics[width=0.58\textwidth]{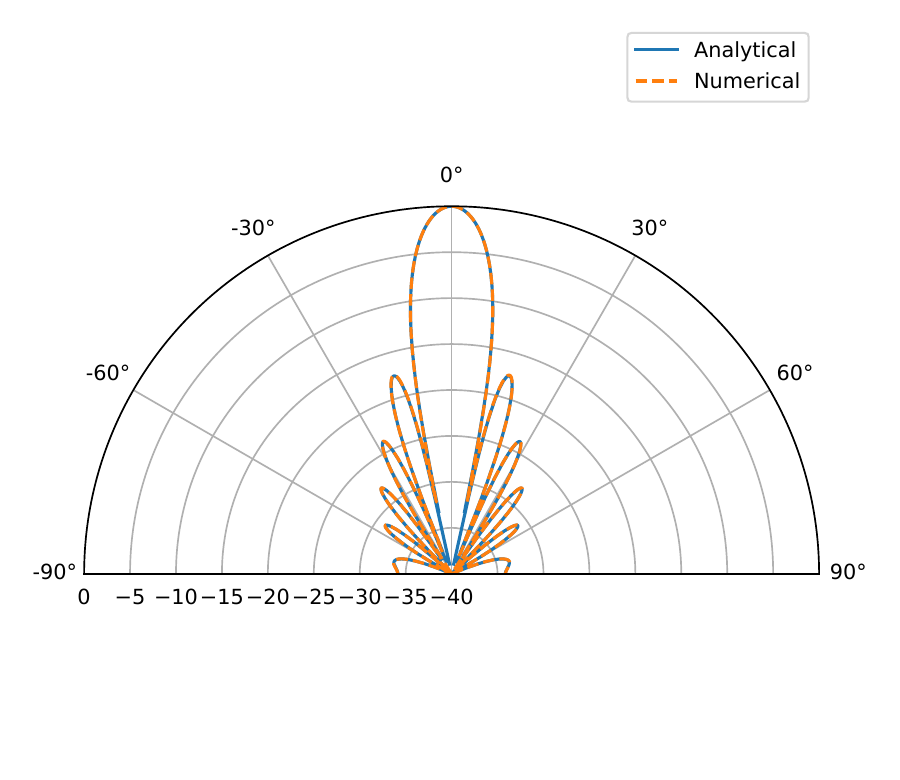}
    \caption{Baffled-piston radiation: far-field directivity pattern at
    $R=4\,\mathrm{m}$.  The analytical solution is compared with the
    reconstructed numerical result on the $N=80$ mesh.  \plotAidDisclosure}
    \label{fig:pistonFarFieldPattern}
\end{figure}

\begin{table}[htbp]
\centering
\begin{tabular}{ccc}
\hline
Cells per wavelength & $h/\lambda$ & $E_{2,|P|}^{\mathrm{ff}}$ \\
\hline
20 & 0.0500 & 2.753e-02 \\
30 & 0.0333 & 1.445e-02 \\
40 & 0.0250 & 1.067e-02 \\
60 & 0.0167 & 8.551e-03 \\
80 & 0.0125 & 8.036e-03 \\
\hline
\end{tabular}
\caption{Mesh-convergence study for the piston-radiation far-field pressure magnitude.}
\label{tab:pistonFarFieldConvergence}
\end{table}

The reconstruction surface lies in the undamped physical domain and meets the
PML onset only at its axial and radial endpoints.  With the piston edge fixed
across all meshes, the near-field error decreases monotonically from
$E_2=2.59\times10^{-2}$ at 20 cells per wavelength to
$2.58\times10^{-3}$ at 80 cells per wavelength.  Over the same sequence,
$E_{2,|P|}^{\mathrm{ff}}$ decreases monotonically from
$2.75\times10^{-2}$ at 20 cells per wavelength to $8.04\times10^{-3}$ at 80
cells per wavelength.  The pressure-magnitude trend therefore provides the
far-field convergence evidence, while
\cref{fig:pistonFarAmpMethod,fig:pistonFarFieldPattern} show that the
reconstructed angular pressure pattern follows the analytical solution.

\FloatBarrier
\subsection{Rigid-sphere radiation force in a standing wave}
\label{sec:gorkovVerification}

The radiation-pressure quantity in \cref{eq:acousticRadiationPressure} is
verified by resolving a sound-hard sphere in a one-dimensional standing wave.
With a rigid reflector at $y=0$, the incident complex pressure and velocity are
\begin{equation}
    P_{\mathrm{inc}}(y)=P_a\cos(ky),
    \qquad
    \V_{\mathrm{inc}}(y)=
    i\frac{P_a}{\rho_0c}\sin(ky)\,\mathbf e_y,
    \label{eq:gorkovStandingWave}
\end{equation}
which is consistent with the $e^{-i\omega t}$ convention in
\cref{as:timeHarmonic}.  For a spherical particle in the Rayleigh regime,
$ka\ll1$, with volume $V_p=4\pi a^3/3$, the Gorkov potential is
\cite{gorkov1962potential}
\begin{equation}
    U_G=V_p\left[
        \frac{f_1|P_{\mathrm{inc}}|^2}{4\rho_0c^2}
        -\frac{3f_2\rho_0|\V_{\mathrm{inc}}|^2}{8}
    \right].
    \label{eq:gorkovPotential}
\end{equation}
Consequently, the axial force is
\begin{equation}
    F_{G,y}=-\frac{\mathrm dU_G}{\mathrm dy}
    =4\pi a^3kE_{\mathrm{ac}}\Phi\sin(2ky),
    \qquad
    E_{\mathrm{ac}}=\frac{P_a^2}{4\rho_0c^2},
    \qquad
    \Phi=\frac{f_1}{3}+\frac{f_2}{2}.
    \label{eq:gorkovForce}
\end{equation}
The sound-hard, acoustically rigid limit has $f_1=f_2=1$ and hence
$\Phi=5/6$.

The numerical force is obtained from the time-averaged radiation-stress
traction on the resolved sphere,
\begin{equation}
    \mathbf F_h=
    \int_{\partial\Omega_p}
    \left[
        p_{\mathrm{rad}}\mathbf I
        +\frac{\rho_0}{2}
        \left(
            \V_{\mathrm{re}}\V_{\mathrm{re}}
            +\V_{\mathrm{im}}\V_{\mathrm{im}}
        \right)
    \right]\cdot\mathbf n_f\,\mathrm dA,
    \label{eq:resolvedRadiationForce}
\end{equation}
where $\mathbf n_f$ points out of the fluid and into the sphere.  The
momentum-flux contribution normal to an exactly sound-hard wall is zero; it is
retained in \cref{eq:resolvedRadiationForce} to state the complete discrete
traction.

\begin{figure}[htbp]
    \centering
    \begin{minipage}[t]{0.58\textwidth}
    \centering
    \resizebox{\linewidth}{!}{%
    \begin{tikzpicture}[x=1cm,y=1cm,>=latex,font=\small]
        \def\domainR{5.2}
        \def\domainH{4.0}
        \def\particleY{2.1}
        \def\particleR{0.45}

        \fill[blue!6] (0,0) rectangle (\domainR,\domainH);
        \draw[very thick] (0,0) -- (\domainR,0);
        \draw[very thick] (\domainR,0) -- (\domainR,\domainH);
        \draw[red!70!black,very thick] (0,\domainH) --
            (\domainR,\domainH);
        \draw[gray!75,dashed,thick] (0,0) -- (0,\domainH);

        \fill[white]
            (0,{\particleY+\particleR})
            arc[start angle=90,end angle=-90,radius=\particleR] -- cycle;
        \draw[black,very thick]
            (0,{\particleY+\particleR})
            arc[start angle=90,end angle=-90,radius=\particleR];
        \fill (0,\particleY) circle (1.4pt);
        \draw[->,thick] (0,\particleY) -- (\particleR,\particleY)
            node[midway,above] {$a$};
        \draw[->,red!70!black,very thick]
            (0.68,{\particleY-0.15}) -- (0.68,{\particleY+0.85})
            node[above] {$F_{h,y}$};

        \draw[<->] (-0.72,0) -- (-0.72,\particleY)
            node[midway,fill=white,inner sep=1pt,rotate=90] {$y_p$};
        \draw[<->] (6.55,0) -- (6.55,\domainH)
            node[midway,fill=white,inner sep=1pt,rotate=90]
            {$H=7.64\,\mathrm{mm}$};
        \draw[<->] (0,-0.62) -- (\domainR,-0.62)
            node[midway,fill=white,inner sep=1pt]
            {$R=10\,\mathrm{mm}$};

        \node[above] at ({0.5*\domainR},\domainH)
            {transducer: $P=P_a\cos(kH)$};
        \node[above] at ({0.5*\domainR},0)
            {rigid reflector: $\partial_nP=0$};
        \node[rotate=90,fill=white,inner sep=1pt] at
            (5.48,{0.5*\domainH})
            {rigid side wall: $\partial_nP=0$};
        \node[rotate=90,fill=white,inner sep=1pt] at (-0.24,3.25)
            {symmetry axis};
        \node[align=left,anchor=west] at (1.05,\particleY)
            {sound-hard sphere\\$\partial_nP=0$};
        \node[anchor=north east,align=right] at
            ({\domainR-0.18},{\domainH-0.18})
            {homogeneous gas\\closed cavity};

        \draw[->,thick] (7.35,0.35) -- (7.35,1.25)
            node[above] {$y$};
        \node at (7.35,0.12) {$\mathbf e_y$};
    \end{tikzpicture}
    }\\[-0.2em]
    \textbf{(a)} Domain and boundary conditions
    \end{minipage}\hfill
    \begin{minipage}[t]{0.38\textwidth}
    \centering
    \includegraphics[width=\linewidth]{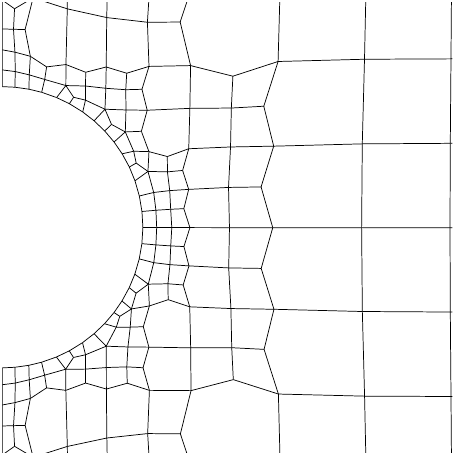}\\[-0.2em]
    \textbf{(b)} Local mesh
    \end{minipage}
    \caption{Rigid-sphere Gorkov case: (a) closed cavity and boundary
    conditions, with the sphere enlarged; (b) illustrative local mesh at
    $h_s/a=0.2$.  Quantitative levels are listed in
    \cref{tab:gorkovMeshConvergence}.  The mesh is a $1^\circ$ axisymmetric
    wedge; no PML is active.  OpenAI Codex (model \texttt{gpt-6-astra}) assisted with the schematic and mesh-rendering code; the mesh is regenerated from the case setup.}
    \label{fig:gorkovDomainSchematic}
\end{figure}
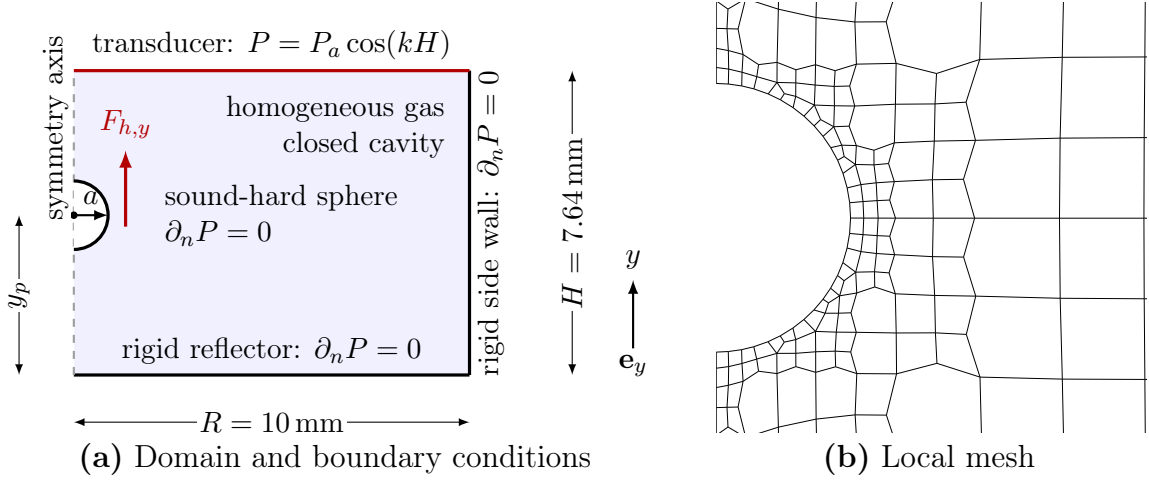

The cylindrical cavity has height $H=7.64\,\mathrm{mm}$ and equal transducer
and reflector radii of $10\,\mathrm{mm}$.  The reflector and cylindrical side
wall are rigid.  At the transducer, the analytical value
$P(H)=P_a\cos(kH)$ is imposed with $P_a=1\,\mathrm{Pa}$,
$f=25.25\,\mathrm{kHz}$, $\rho_0=1.2\,\mathrm{kg\,m^{-3}}$, and
$c=343\,\mathrm{m\,s^{-1}}$.  Thus, only the excitation is prescribed; the
total pressure and scattered field around the sphere are computed.  No PML is
active in this closed-cavity reference problem.  The computational domain and
boundary conditions are summarized in \cref{fig:gorkovDomainSchematic}.

\begin{figure}[htbp]
    \centering
    \includegraphics[width=0.92\textwidth]{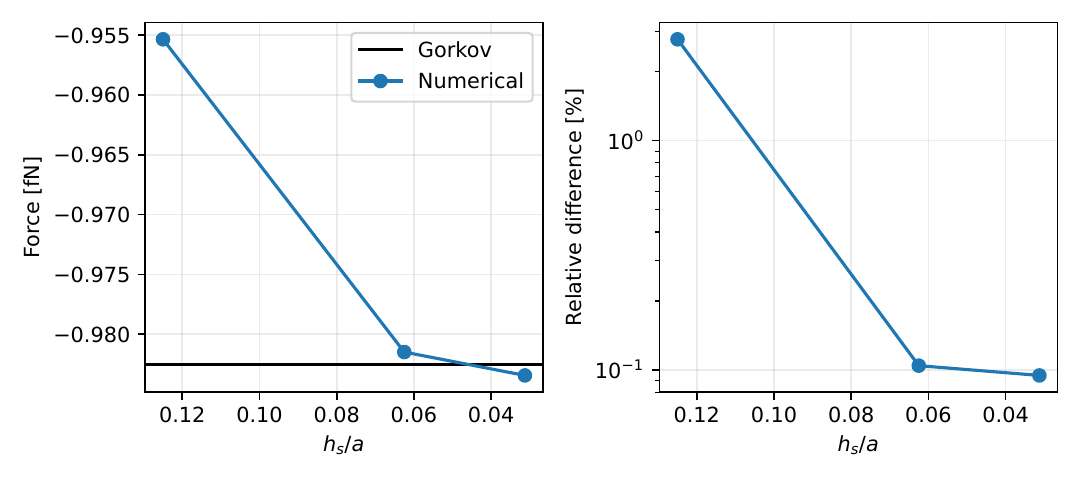}
    \caption{Rigid-sphere force sensitivity to the near-sphere cell size and
    surface segmentation.  \plotAidDisclosure}
    \label{fig:gorkovMeshConvergence}
\end{figure}

The axisymmetric unstructured meshes contain approximately $2.2\times10^4$
cells, with $35^\circ$--$42^\circ$ maximum and less than $2.1^\circ$ average
non-orthogonality.  The prescribed local cell size adjacent to the sphere is
denoted by $h_s$, and $h_s/a$ is the dimensionless near-sphere resolution.
\Cref{tab:gorkovMeshConvergence} and
\cref{fig:gorkovMeshConvergence} show the near-sphere refinement study at
$a=60\,\mu\mathrm{m}$ and $y=4.0003\,\mathrm{mm}$.  Refining from
$h_s/a=1/8$ to $1/16$ changes the force by $2.665\%$; the change from $1/16$ to
$1/32$ is $0.199\%$.  The medium and fine results differ from the Gorkov value
by $0.104\%$ and $0.095\%$, respectively.  The medium level is therefore used
for the parameter sweeps.

\begin{table}[htbp]
    \centering
    \caption{Near-sphere mesh sensitivity for the rigid sphere with
    $a=60\,\mu\mathrm{m}$.  The Gorkov reference is
    $F_{G,y}=-0.982526\,\mathrm{fN}$ and
    $E_F=|F_h-F_G|/|F_G|$.}
    \label{tab:gorkovMeshConvergence}
    \small
    \begin{tabular}{lrrrrr}
        \toprule
        Level & $h_s/a$ & Segments & Faces & $F_{h,y}$ (fN) & $E_F$ (\%) \\
        \midrule
        Coarse & 0.12500 & 64  & 67  & -0.955343 & 2.767 \\
        Medium & 0.06250 & 128 & 131 & -0.981500 & 0.104 \\
        Fine   & 0.03125 & 256 & 259 & -0.983455 & 0.095 \\
        \bottomrule
    \end{tabular}
\end{table}

\begin{figure}[htbp]
    \centering
    \includegraphics[width=0.82\textwidth]{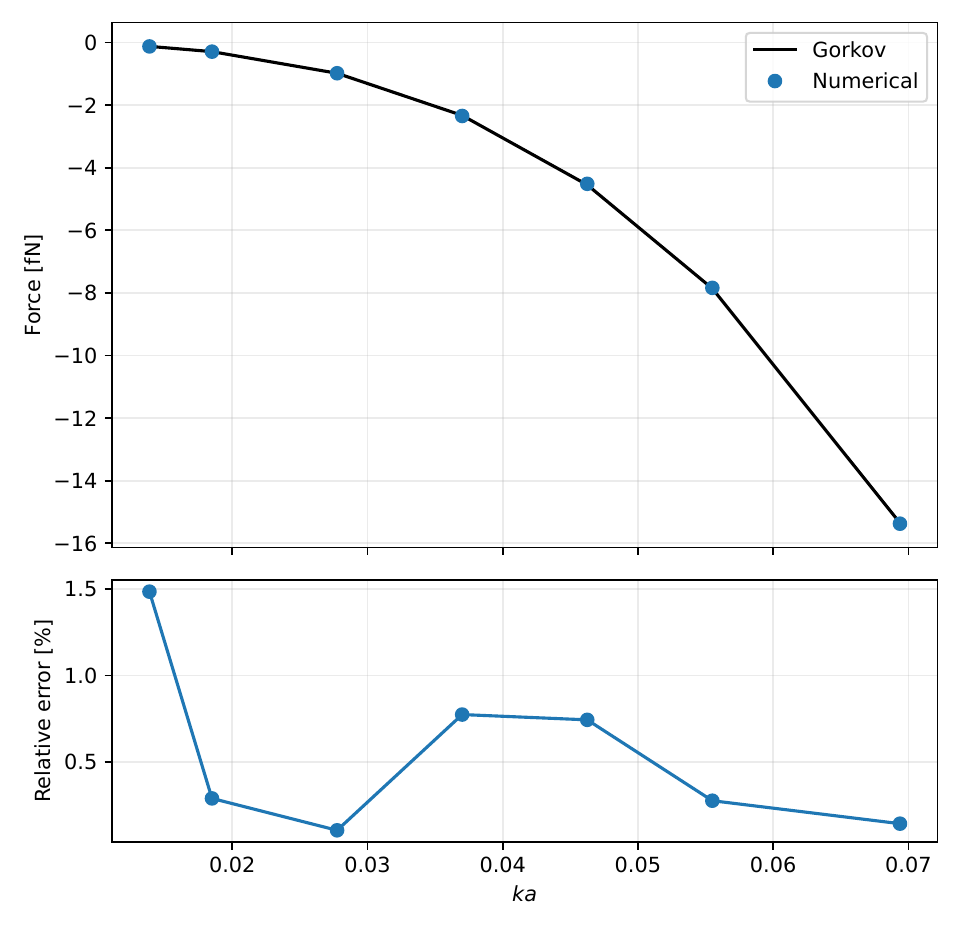}
    \caption{Resolved rigid-sphere force and Gorkov prediction over the
    Rayleigh size-parameter range.  \plotAidDisclosure}
    \label{fig:gorkovRadiusSweep}
\end{figure}

The radius sweep covers $30\leq a\leq150\,\mu\mathrm{m}$, corresponding to
$0.0139\leq ka\leq0.0694$.  The numerical force follows the expected
$a^3$ scaling, and the maximum difference from \cref{eq:gorkovForce} is
$1.485\%$, as shown in \cref{fig:gorkovRadiusSweep}.  This range remains in the
Rayleigh regime required by the point-particle reference.

\begin{figure}[htbp]
    \centering
    \includegraphics[width=0.88\textwidth]{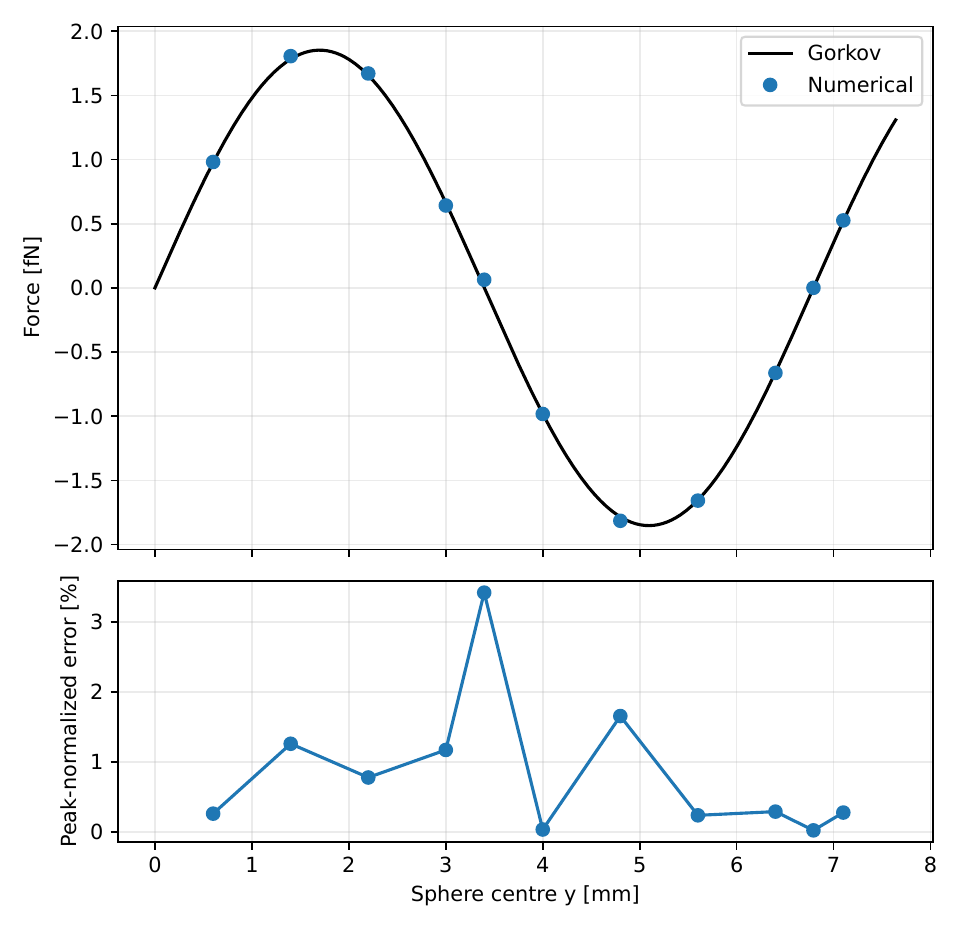}
    \caption{Axial radiation force on a $60\,\mu\mathrm{m}$ rigid sphere as a
    function of its vertical position.  The error is normalized by the peak
    analytical force so that the zero-force positions remain well defined.  \plotAidDisclosure}
    \label{fig:gorkovPositionSweep}
\end{figure}

Finally, a $60\,\mu\mathrm{m}$ sphere is moved through the standing wave.  The
resolved force reproduces both force lobes, their signs, and the two analytical
zero crossings in the cavity, as shown in \cref{fig:gorkovPositionSweep}.  The
root-mean-square force difference is $1.288\%$ of the peak Gorkov force.  The
maximum peak-normalized difference is $3.415\%$ at the first analytical zero;
there the numerical residual is $0.0633\,\mathrm{fN}$.  At an analytical
zero, this residual is assessed using peak-force normalization.  The present
comparison does not separate spatial-discretization effects from differences
between the resolved finite sphere in the bounded cavity and the point-particle
reference.

\section{Discussion}
\label{sec:discussion}

The material treatment in \cref{sec:materialRepresentation} uses
volume-averaged density and compressibility in cells and the reconstructed
phase fraction averaged over the available adjacent PLIC cuts for the face
density.  This definition is independent of the velocity direction and supplies
a common scalar fraction on processor faces.  The controlled comparisons in
\cref{sec:geometricComparison,tab:geometricComparison} show essentially
unchanged pressure errors on the tested planar and spherical cases, while
reducing the observed decomposition dependence to roundoff.  This is evidence of reproducibility, not of a
general improvement in pressure or force accuracy.  The resulting inverse face density is a
one-field closure, not an exact sharp cut-interface quadrature.  The layered
case assesses this closure for the stated interface placements; uniformly
second-order accuracy for arbitrary subcell interface positions is not claimed,
and specialized sharp-interface flux treatments are outside the present scope.

The PML coefficients listed in the mathematical model are local functions of
the damping tensor and the acoustic reaction coefficient.  This keeps the
real-valued PML equations compatible with heterogeneous media and allows the
same unstructured FVM block structure in the physical domain and absorbing layer.

The homogeneous baseline gives approximately second-order pressure convergence
on orthogonal meshes and on meshes with a non-orthogonal interior and one
orthogonal boundary-cell layer.  This verifies the corrected interior flux
discretization separately from boundary-condition errors.  Reconstructed
velocity converges more slowly on the non-orthogonal interior mesh family, and the
boundary-skewed diagnostic loses pressure order.  Consequently, no general
second-order claim is made for arbitrary polyhedral meshes or arbitrary
boundary-cell geometry.

The layered reference verifies the complete complex pressure across the gas,
liquid, and PML with one global error measure.  Its damping study shows the
expected loss of accuracy when outgoing waves are not sufficiently attenuated
before the outer boundary, while the aligned refinement sequence demonstrates
monotonic convergence of the coupled interface--PML solution.  For the piston,
the mesh sequence demonstrates convergence of the reconstructed far-field
pressure magnitude, while the far-field sound-pressure-level comparison shows
the corresponding angular radiation pattern.

The rigid-sphere study directly verifies the radiation-stress postprocessing.
The radius sweep recovers the Rayleigh $a^3$ scaling, while the position sweep
recovers the spatial force phase and sign changes.  The sub-percent
medium-to-fine force change and the small momentum-flux contribution at the
sound-hard surface show that the integrated result is controlled by the
radiation-pressure field rather than by an unresolved normal-velocity term.

\section{Conclusions}
\label{sec:conclusions}

This work presents a Helmholtz--PML finite-volume method on unstructured meshes
for frequency-domain acoustics in heterogeneous two-phase media.  It combines
a volume-fraction material representation with a real-valued split PML and a
block system for the real and imaginary pressure components.  Geometric face
averaging provides a velocity-independent coefficient while retaining the
established acoustic flux, reaction quadrature and pressure-gradient
postprocessing.  The
homogeneous baseline gives second-order pressure convergence on orthogonal and
non-orthogonal interior meshes, with finest orders of $2.04$ and $1.94$ for the
Dirichlet and mixed cases, respectively.  Velocity converges at approximately
second order on the orthogonal family and at about order $1.7$ on the
non-orthogonal interior mesh family.  In the one-dimensional layered case, the numerical
and analytical complex pressures are compared over the complete gas--liquid--PML
domain, giving $E_P=1.197\times10^{-3}$ at $N_x=8000$.  With the interface and
PML start aligned during refinement, $E_P$ decreases monotonically to
$2.657\times10^{-4}$ at $N_x=8960$.

The baffled piston confirms multidimensional radiation, with a $0.258\%$
near-field error and a $0.804\%$ far-field pressure-magnitude error on the
finest mesh.  The rigid-sphere
force differs from the Gorkov prediction by at most
$1.485\%$ over the tested Rayleigh size range.  Its peak-normalized RMS
difference over the position sweep is $1.288\%$.

Future work should extend the acoustic model to include bulk attenuation due
to viscosity and dissipative losses at solid walls.  The latter
may be represented by resolving the thermoviscous boundary layers or by an
appropriate effective wall condition.  A validated treatment of these loss
mechanisms would provide the first-order acoustic field required for a
subsequent coupling to a second-order, time-averaged flow model for acoustic
streaming.  A further extension could couple the bulk acoustic field to an
interface-displacement equation through the linearized Young--Laplace
condition, thereby including first-order capillary--acoustic interaction.

\begin{samepage}
\section*{Code and data availability}

The layered-interface plots and sensitivity data use the selected geometric
method and have been checked against the original implementation as described
in \cref{sec:geometricComparison}.  The homogeneous-wave and piston results
are retained from the single-phase benchmark runs, using the archived
cell-volume-weighted convergence data and the geometry-aligned piston
refinements, respectively.  The rigid-sphere force curves are reprocessed
from the archived integrated forces.  In all three single-phase benchmarks,
the acoustic domain contains only gas: the rigid sphere is a sound-hard
boundary, not a liquid inclusion.  Both area definitions therefore give
$\alpha_f=0$ and the same acoustic coefficients; these benchmarks do not
exercise PLIC averaging.  The accompanying provenance record identifies
figure and table inputs and distinguishes archived results from the new
two-phase comparisons.

The implementation and case workflows are maintained in the
\texttt{TwoPhaseFlow} repository,
\url{https://github.com/tmaric/TwoPhaseFlow}, on the development branch
\texttt{feature/acousticLevitation}.  The case dictionaries associated with
this revision explicitly select the following entries in the
\texttt{acousticInterface} subdictionary of \texttt{system/fvSchemes}:
\begin{verbatim}
acousticInterface
{
    areaFraction       plicAverage;
    flux               legacy;
    writeDiagnostics   false;
}
\end{verbatim}
\end{samepage}
Here \texttt{legacy} identifies the established flux in
\cref{eq:retainedInterfaceFlux}; it does not select the former face-area
calculation.  The latter remains available through
\texttt{areaFraction legacy}.  Missing entries retain the original behavior
for compatibility, so the submission cases specify both selections.
The case mapping and verification manifests are documented in
\texttt{SUBMISSION\_ACOUSTICS.md} in the repository.  A branch name alone is
not an immutable source identifier.  The accompanying implementation is
available at commit \href{https://github.com/tmaric/TwoPhaseFlow/tree/bb09dbfb34c1830478feecd8dd445957ac15949c}{\texttt{bb09dbfb34c1}}~\cite{XuLiuMaric2026AcousticSoftware};
the manifests distinguish the source and binary hashes used by each study.

All secondary data used in the figures, tables and supporting comparisons
are collected in \path{data/publications/helmholtz-pml/} in the same
repository.  The package contains sampled pressure profiles, convergence
metrics, integrated forces, analytical reference values, data dictionaries
and scripts that regenerate the numerical plots and tables without
OpenFOAM.  The root \texttt{PROVENANCE.md} maps every figure and table to
its repository-relative data source.  A checksum manifest identifies the
exact package contents.  Raw simulation fields and meshes are not included
in this secondary-data package.

Version 1.0 of the secondary-data archive is deposited in
TUdatalib~\cite{XuMaric2026HelmholtzPMLData}.  The dataset has the reserved
DOI \url{https://doi.org/10.48328/tudatalib-2357}; DOI registration and public
release remain pending.  The deposited files are identified by the checksum
manifest and the deposit record linked from the repository provenance.

\section*{Acknowledgements}

The authors gratefully acknowledge the computing time provided on the
high-performance computer Lichtenberg II at TU Darmstadt, funded by the
German Federal Ministry of Research, Technology and Space (BMFTR), and the
State of Hesse Ministry of Science and Research, Art and Culture (HMWK).

\section*{Funding}

Mario Kupnik and Dieter Bothe acknowledge support through their tandem
project, funded by the Deutsche Forschungsgemeinschaft (DFG, German Research
Foundation) -- project number 542327521.
The funder had no role in study design; data collection, analysis or
interpretation; manuscript preparation; or the decision to submit the article
for publication.

\section*{Declaration of competing interest}

The authors declare that they have no competing financial interests or
personal relationships that could have influenced the work reported in this
paper.

\section*{CRediT authorship contribution statement}

\noindent\textbf{Chuanchao Xu}: Data curation, Formal analysis, Investigation, Methodology, Software, Visualization, Writing -- original draft.

\noindent\textbf{Jun Liu}: Formal analysis, Investigation, Methodology, Software.

\noindent\textbf{Jan Helge Dörsam}: Validation, Writing -- review \& editing.

\noindent\textbf{Mario Kupnik}: Conceptualization, Funding acquisition, Project administration, Validation, Writing -- review \& editing.

\noindent\textbf{Tomislav Maric}: Conceptualization, Data curation, Formal analysis, Investigation, Methodology, Software, Supervision, Writing -- review \& editing.

\noindent\textbf{Dieter Bothe}: Conceptualization, Formal analysis, Methodology, Project administration, Resources, Supervision, Writing -- review \& editing.

\section*{Declaration of generative AI and AI-assisted technologies in the manuscript preparation process}

During revision of this work, OpenAI Codex was used under the direction of
the corresponding author to assist with manuscript review and revision,
software development, verification and data-processing scripts, and
preparation of reproducible visualizations and TikZ schematics.  The
authors retain responsibility for the scientific content, interpretation,
and final manuscript.

\bibliographystyle{elsarticle-num}
\paperbibliography{bibliography}
\end{document}